\documentclass[sigconf]{acmart}
\usepackage{todonotes}
\usepackage{caption}
\usepackage{subcaption}
\usepackage{multicol}
\def\BarGrey#1{{\color{black!70}\rule{#1mm}{8pt}}}

\AtBeginDocument{%
  \providecommand\BibTeX{{%
    \normalfont B\kern-0.5em{\scshape i\kern-0.25em b}\kern-0.8em\TeX}}}

\copyrightyear{2023}
\acmYear{2023}
\setcopyright{acmlicensed}\acmConference[CHI '23]{Proceedings of the 2023 CHI Conference on Human Factors in Computing Systems}{April 23--28, 2023}{Hamburg, Germany}
\acmBooktitle{Proceedings of the 2023 CHI Conference on Human Factors in Computing Systems (CHI '23), April 23--28, 2023, Hamburg, Germany}
\acmPrice{15.00}
\acmDOI{10.1145/3544548.3580990}
\acmISBN{978-1-4503-9421-5/23/04}
\begin{document}

\title[Reciprocity, Homophily, and Social Network Effects in Pictorial Communication]{Reciprocity, Homophily, and Social Network Effects in Pictorial Communication: A Case Study of Bitmoji Stickers}
\author{Julie Jiang}

\authornote{Work was done when the authors were interns at Snap Inc.}
\email{juliej@isi.edu}
\affiliation{%
  \institution{University of Southern California}
  \state{CA}
  \country{{USA}}
 }
\author{Ron Dotsch}
\email{rdotsch@snap.com}
\affiliation{\institution{Snap Inc.} \state{CA} \country{USA}}
\author{Mireia Triguero Roura}
\authornotemark[1]
\email{mt3197@columbia.edu}
\affiliation{
    \institution{Columbia University}
    \state{NY}
    \country{USA}}
\author{Yozen Liu}
\email{yliu2@snap.com}
\affiliation{\institution{Snap Inc.} \state{CA} \country{USA}}
\author{Vitor Silva}
\email{vitor.silva.sousa@gmail.com}
\affiliation{\institution{Snap Inc.} \state{CA} \country{USA}}
\author{Maarten W. Bos}
\email{maarten@snap.com}
\affiliation{\institution{Snap Inc.} \state{CA} \country{USA}}
\author{Francesco Barbieri}
\email{fbarbieri@snap.com}
\affiliation{\institution{Snap Inc.} \state{CA} \country{USA}}
\renewcommand{\shortauthors}{}

\begin{abstract}
Pictorial emojis and stickers are commonly used in online social communications. We analyzed social communications using Bitmoji stickers, which are expressive pictorial stickers made from avatars resembling actual users. We collect a large-scale dataset of 3 billion Bitmoji stickers' metadata, shared among 300 million Snapchat users. We find that individual Bitmoji sticker usage patterns can be characterized jointly on dimensions of reciprocity and selectivity. Generally speaking, users are either both reciprocal and selective about whom they use Bitmoji stickers with or neither reciprocal nor selective. We additionally demonstrate network homophily by showing that friends use Bitmoji stickers at similar rates. Finally, using a quasi-experimental approach, we show that receiving Bitmoji stickers from a friend encourages future Bitmoji sticker usage and overall Snapchat engagement. Our work carries implications for a better understanding of online pictorial communication behaviors.
\end{abstract}


\begin{teaserfigure}
  \includegraphics[width=\textwidth]{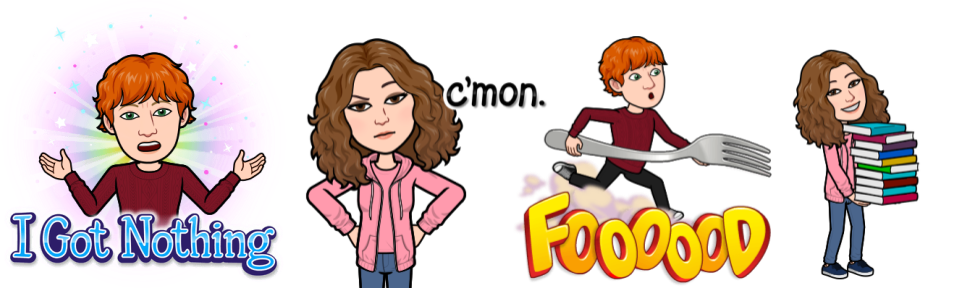}
  \caption{Bitmoji stickers made with avatars that resemble Ron Weasley and Hermione Granger.}
  \label{fig:teaser}
\end{teaserfigure}

\maketitle

\section{Introduction}














Pictorial methods of virtual communications, including images, videos, emoticons, emojis, stickers, GIFs, and memes, have amassed considerable adoption in computer-mediated communications over the years \cite{vox2015}. Research shows that such non-verbal communications convey attitudes and enhance emotion \cite{lo2008nonverbal,tang2019emoticon,tigwell2016oh}. They also increase the sender's perceived levels of intimacy, positivity, and social connectedness \cite{utz2000social, tung2007increasing,byron2007mail,skovholt2014communicative,janssen2014affective,tigwell2016oh,gesselman2019worth,shandilya2022need}. It has also been shown that using pictorial communication establishes interpersonal bonds \cite{utz2000social,shandilya2022need}. Pictorial communication methods complement textual communication with non-verbal cues, similar to the way things like body language, eye contact, micro-expressions, and tone complement in-person verbal communication \cite{argyle1971communication,lo2008nonverbal}. 

Despite the importance and prolific use of pictorial features in online communication, less is known about the characteristics of pictorial communication behavior within social networks. We aim to bridge this gap in research by contextualizing a user's pictorial communication frequencies with those of their friends in the social network in a relatively controlled environment. We approach our goal through a case study of Bitmoji sticker usage on Snapchat, a messaging app where users can send Bitmoji stickers to each other. Bitmojis are visual avatars of the users created by the users themselves, which is why they often reflect how users look physically. The stickers are then rendered from templates curated and designed in-house by the Bitmoji developers. These Bitmoji stickers are capable of displaying a range of emotional expressions and portraying the users engaging in a variety of activities, making them highly expressive and appealing to use. Fig. \ref{fig:teaser} shows examples of the Bitmoji stickers based on avatars that --- for illustrative purposes --- were made to look like Ron Weasley and Hermione Granger, two characters from the Harry Potter franchise. 

In this work, we use Bitmoji as an example of a type of pictorial language that is socially constructed and spread to develop insights as to how communication behaviors form and evolve. Bitmoji is functionally similar to other pictorial communication, making it an appropriate pictorial communication method to study. Bitmojis might be preferable to users because they are unique enough so that users have reasons to adopt them over existing methods such as emojis, memes, or GIFs --- Bitmoji stickers can be customized by each user to uniquely resemble themselves physically. Finally, Bitmoji support is built into Snapchat. 
The same cannot be said for emojis, memes, GIFs, etc. since they can easily transfer from platform to platform \cite{jiang2018perfect}. By tracking Bitmoji sticker usage data on Snapchat, we can comprehensively characterize user behavior and study changes in those behaviors, rendering a pseudo-controlled environment in which we can study one type of pictorial communication method in isolation.

We utilize a large-scale dataset of 3 billion Bitmoji sticker usage metadata among 289 million distinct Snapchat users throughout the month of May 2022. The scope of this work is limited to whether and when any type of Bitmoji sticker is shared between two friends. We explicitly do not study the content of the Bitmoji stickers users send. We carry out our analyses from three perspectives of communication behavior: \underline{reciprocity/selectivity}, \underline{network homophily}, and \underline{social contagion}. Drawing on related literature on reciprocity, we hypothesize that the behavior could be reciprocal, with some users matching the level of Bitmoji stickers sent to them by their friends. Users' Bitmoji sticker usage behavior could also be characterized by how selective they are in using the stickers with their friends. A highly selective user only uses Bitmoji stickers with a few friends, whereas a highly unselective user will use them with all of their friends indiscriminately. While these two concepts --- reciprocity and selectivity --- gauge the behavior differently, we argue and show that they are highly correlated. If some users are reciprocal regarding Bitmoji sticker usage, that would mean this behavior can be contextualized in the presence of the social network. A logical next step is to examine whether this behavior reflects network homophily \cite{mcpherson2001birds}, which would suggest that users use Bitmoji stickers at similar frequencies as their friends. Leveraging quasi-experimental methods, we additionally investigate if this behavior is socially contagious. That is, users are more likely to use Bitmoji stickers after seeing their friends use them. We further analyze whether this contagion extends not only to Bitmoji stickers but also in-app engagement overall.

Our main contributions are summarized as follows:
\begin{enumerate}
    \item On using (any) Bitmoji stickers, users range from being reciprocal and selective to being nonreciprocal and unselective. Reciprocity and selectivity are two distinct but correlated measures of activity.
    \item Bitmoji sticker usage frequencies exhibit network homophily--users tend to use (any) Bitmoji stickers at the same rate as their friends.
    \item Receiving any Bitmoji sticker encourages all forms of future Bitmoji sticker usage and overall in-app engagement, and this effect is most pronounced for reciprocal and selective users.
\end{enumerate}

Through this case study, we believe our work facilitates a deep understanding of social processes underpinning pictorial communication styles on social networks. We discuss the broader implications of our findings for the HCI community and social media platform designers.

\section{Background and Related Work}

\subsection{Bitmoji Stickers and Snapchat}

Bitmoji is a social media app where users create avatars of themselves \cite{digital2021bitmoji}. This avatar is a highly customizable cartoon-like rendition of the user, complete with hundreds of options for hairstyle, hair color, glasses, headwear, makeup, outfits, and more \cite{bitmojideluxe}. Though Bitmoji is a standalone app that can support other third-party services, it is owned by Snap Inc., the company that also makes Snapchat, and is natively integrated with Snapchat. At 323 million users as of the first quarter of 2022, Snapchat is a popular camera and instant messaging app known for the ephemerality of its content \cite{snap2022q1}. 

Though not required, the vast majority of Snapchat users have a Bitmoji avatar. Over 1 billion Bitmoji avatars have been created as of mid-2022 \cite{techcrunch2022}. Snapchat users with Bitmoji avatars can send highly expressive and entertaining Bitmoji stickers that resemble them to their friends on Snapchat (e.g., Fig. \ref{fig:teaser}). They can be sent either as photos or videos `Snaps' or textual chats. These Bitmoji stickers are created from filling the user's self-designed avatars in pre-designed templates created by the Bitmoji developers at Snap Inc. Users can also send Friendmojis to each other, which are two-person Bitmoji stickers with both friends' Bitmoji avatars.

Stickers are common features present in many social media apps beyond Snapchat, including Instagram, Wechat, and LINE \cite{tang2019emoticon}. \citet{de2018biaoqing} defined stickers as collections of images similar to graphical emoticons and emojis that are curated and/or personalized on instant messaging and social media apps. 
Some stickers are curated and created by the platform, others are custom-made by users through uploading a picture of their choosing. Bitmoji stickers lie somewhere in the middle. The templates are pre-desgined, but the avatars are created by users. Users can also create their own Bitmoji stickers with custom embedded text \cite{bitmojicustomize}. 

Compared to other forms of pictorial communication, Bitmoji sticker stand out in two aspects: functionally for users and analytically for researchers. Functionally, Bitmoji stickers conveniently balance high-quality, pre-designed stickers with the uniqueness of individually curated avatars, making them more engaging and special. These are reasons users might choose Bitmoji stickers over the one-size-fits-all alternatives of emojis, GIFs, or memes. Analytically, studying Bitmoji stickers on Snapchat affords researchers a unique opportunity to study pictorial communication on a single platform. Almost all other pictorial communication methods are transferable cross-platform, therefore any social network effects or temporal developments of usage characteristics cannot be accurately measured given access to only one platform. Bitmoji stickers, however, are easily accessible and primarily used on Snapchat, allowing us to study a single mode of pictorial communication in a relatively controlled environment.

\subsection{Reciprocity and Selectivity in Social Networks}

There is substantial literature on reciprocity, the concept that people return behaviors or exchange things, either altruistically or motivated by mutual benefit \cite{gintis2000strong,fehr2000fairness,smith2014social,bisberg2022gift}. In social science literature, researchers further explored the theory of the \textit{reciprocity of liking} \cite{eastwick2009reciprocity,eastwick2007selective}, which postulates that people are more likely to be attracted to those who they perceive to be attracted to them, thereby reciprocating affection. However, not all reciprocity of liking is created equal. In a speed dating experiment, \citet{eastwick2007selective} found that participants are more attracted to people who are selectively attracted to themselves and less attracted to people who are generally attracted to everyone. 

That leads us to the next topic: selectivity in social behavior. In interpersonal relationships, humans are drawn to some people, but not everyone. This applies both in romantic situations and friendships. This could be due to a wide and complex range of factors -- similarity in interests or values, physical attractiveness, impressions, compatibility, reciprocity, and environmental factors \cite{byrne1973interpersonal,duck1973personal,huston1978interpersonal,hallinan1978process,eastwick2009reciprocity}. The socioemotional selectivity theory also posits that it could be related to age \cite{carstensen1999taking}. As humans age, their perception of their lifetime horizon reduces, and thus they tend to reprioritize their goals and activities, opting to deepen close relationships and become more socially selective \cite{carstensen1999taking}.
\begin{figure*}
    \centering
    \includegraphics[width=0.9\textwidth]{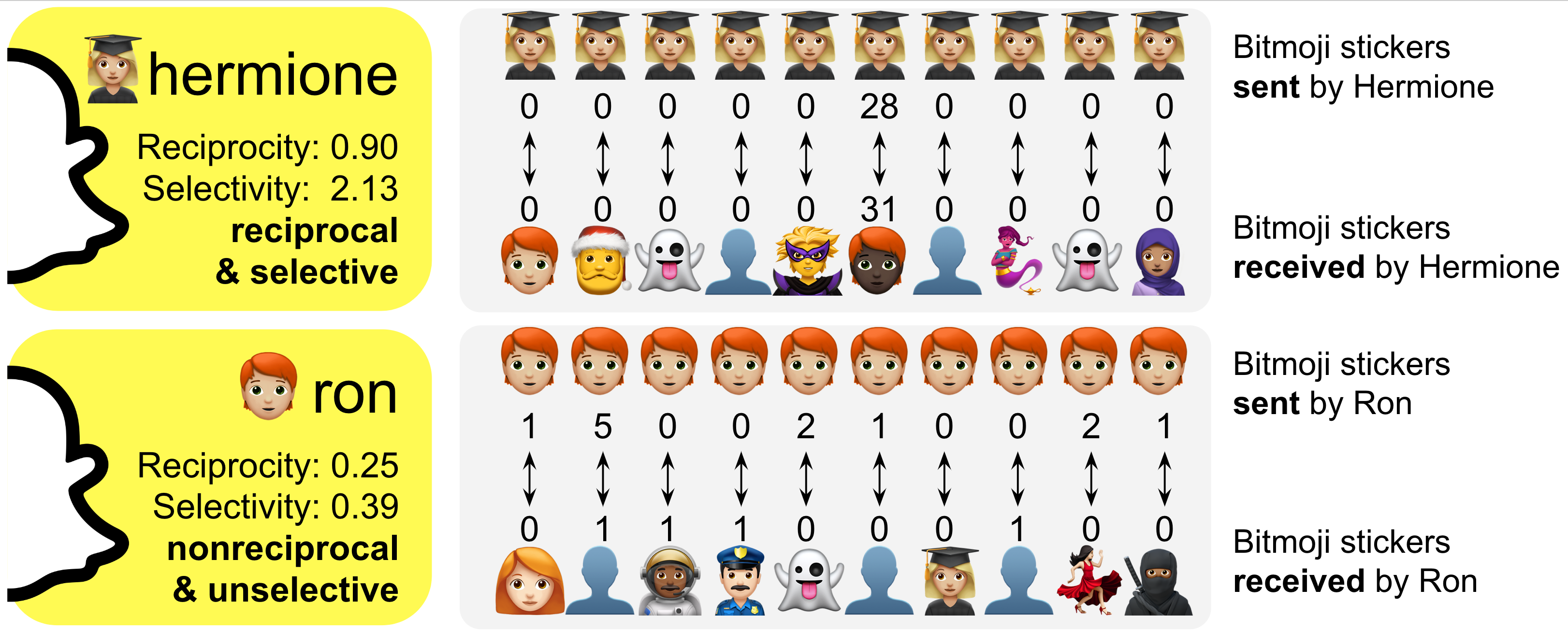}
    \caption{Toy examples of two hypothetical, quintessential frequent Bitmoji sticker users. Hermione is a reciprocal and selective user of Bitmoji stickers, whereas Ron is a nonreciprocal and unselective Bitmoji sticker user.}
    \label{fig:reciprocity_selectivity_toy}
\end{figure*}

In this work, we characterize users on dimensions of reciprocity and selectivity by measuring how frequently friends use Bitmoji stickers with each other. In terms of reciprocity, a reciprocal Bitmoji sticker user matches the amount of Bitmoji stickers they received from their friends, whereas a nonreciprocal user sends Bitmoji stickers irrespective of how many they receive in return. In terms of selectivity, a user can either selectively only use Bitmoji stickers with a few friends or unselectively use Bitmoji sticker with all of their friends. An unselective user could be someone who enjoys using Bitmoji stickers under any context in any type of Snapchat conversation.  We discuss how we quantify reciprocity and selectivity in \S\ref{sec:measure_reciprocity_selectivity}. 

While these two measures of user behavior --- reciprocity and selectivity --- are distinct, we hypothesize that they would be actually correlated. We illustrate this with two toy examples in Fig. \ref{fig:reciprocity_selectivity_toy}. Both Ron and Hermione are frequent Bitmoji sticker users because they send them quite frequently in chats, but that is where their commonality ends. Hermione is a highly reciprocal and selective Bitmoji sticker user, sending lots of Bitmoji stickers to and only to one friend, who reciprocates nearly an equivalent amount of Bitmoji stickers. On the other hand, Ron is a nonreciprocal and unselective Bitmoji sticker user in that he sends a few Bitmoji stickers, regardless of reciprocity, to lots of users. Putting our hypothesis to the test, we raise our first research question:

\setlength\fboxsep{3mm}
\medskip
\noindent\fbox{%
    \parbox{
    \dimexpr\linewidth-2
    \fboxsep-2
    \fboxrule
    }{%
        \textbf{RQ1:} How can we characterize Bitmoji sticker usage frequency on dimensions of reciprocity and selectivity?
    }%
}\medskip

\subsection{Homophily and Contagion on Social Networks}
Homophily is a social network theory that postulates that users who are in similar groups, communities, and clusters share similar characteristics, behaviors, and interests \cite{mcpherson2001birds,kossinets2009origins,bisgin2012study,figeac2021behavioral}. Homophily can be observed in close, intimate relationships and in mere acquaintances \cite{mcpherson2001birds}. They can manifest as similarities in race, gender, age, religion, education, and social class \cite{mcpherson2001birds}. Additionally, human behavior such as academic achievement, marijuana use, sexual activities, and political orientations were all shown to be homophilous in real-world friendships \cite{cohen1977sources,kandel1978homophily,billy1984adolescent,knoke1990networks}. \citet{centola2011experimental} found that the adoption of health behaviors was more prominent in homophilous networks. On social media, users who share similar political or entertainment content feel closer to one another \cite{figeac2021behavioral}. In this work, we analyze the homophilous behavior of sending Bitmoji stickers. Specifically, we question if users who are friends with each other use Bitmoji stickers at similar frequencies:

\medskip
\noindent\fbox{%
    \parbox{
    \dimexpr\linewidth-2
    \fboxsep-2
    \fboxrule
    }{%
        \textbf{RQ2:} Is the frequency of Bitmoji stickers usage homophilous?
    }%
}\medskip

Closely related to network homophily is the social contagion phenomenon, which argues that one person's behaviors and emotions can socially influence another person to whom they are connected in the social network \cite{rosenquist2011social,pachucki2011social,christakis2013social,kramer2014experimental,tsvetkova2014social,hodas2014simple,ferrara2015measuring,ferrara2015quantifying,monsted2017evidence}. The theory of social contagion offers one explanation as to why we observe homophilous behaviors in social networks. Take for example the context of this work. If users are found to use Bitmoji stickers at similar frequencies, the theory of social contagion suggests that one user's use of Bitmoji stickers will influence their friend's decision to use Bitmoji stickers. Another explanation of behavioral homophily is that some latent homophilous attributes can influence both two users' desire to use Bitmoji stickers and their friendship formation. For instance, the shared interests in pictorial communication of two users led them to both use Bitmoji stickers and become friends on Snapchat. In observational studies such as ours, it is unfortunately extremely challenging, if not impossible, to disentangle the effects of pure contagion from the effects of pure homophily \cite{shalizi2011homophily}. This is due to our inability to control for unobservable covariates that result both in two users sharing network ties \textit{and} adopting similar behavior. 

In this study, we aim to measure the effects of receiving Bitmoji stickers on users' future behavior. In an ideal world, we would control for latent homophily and measure the effects of the social contagion of using Bitmoji stickers using causal inference approaches. For example, we can subject users to repeated randomized controlled trials, randomly manipulating each user to either receive the treatment (which is receiving Bitmoji stickers) or not. In the context of platform-specific studies, this can be done with A/B testing \cite{kohavi2017online}. However, there are drawbacks to such approaches, one being the time, resources, and infrastructure support required to conduct such large-scale experiments \cite{hariton2018randomised}. More importantly, a study in which we would manipulate messaging content in private communication faces serious ethical, privacy, and legal consequences if users were not adequately informed. However, if users were informed, the results may be jeopardized by the observer or Hawthorne effect \cite{mccarney2007hawthorne}, which is an individual modifying their behavior because they are aware they are being observed. 

Given these limitations, we follow prior social media observational studies \cite{de2016discovering,olteanu2017distilling,saha2019social,maldeniya2020herding,saha2021advertiming,bisberg2022gift} and utilize propensity score matching, a quasi-experimental approach suitable for observational studies \cite{rosenbaum1984reducing,gelman2007causal,imbens2015causal}. Though we recognize the limitations of our quasi-experimental settings and cannot conclude any true causality, this quasi-experimental method minimizes confounders induced by observable covariates and is thus preferable to direct correlational analyses \cite{imbens2015causal}. Specifically, we measure how receiving Bitmoji stickers influences future Bitmoji stickers usage. Further, since users might find the platform more interesting upon receiving Bitmoji stickers, we examine if user engagement will rise afterwards. We phrase our final research question as follows:

\medskip
\noindent\fbox{%
    \parbox{
    \dimexpr\linewidth-2
    \fboxsep-2
    \fboxrule
    }{%
        \textbf{RQ3:} Does receiving Bitmoji stickers encourage future Bitmoji sticker usage and overall in-app engagement?
    }%
}\medskip

\section{Data}\label{sec:data}

We leverage a large-scale Snapchat dataset collected from May 1, 2022, to May 31, 2022. We limit our study to users who are over the age of 18 and who have Bitmoji avatars. In total, we collected  correspondence metadata and bidirectional friendships among 289 million unique users. Our correspondence data collection is only limited to whether each correspondence contains a Bitmoji sticker or not, as well as the time and involved users. We do not collect any additional private correspondence information or message details. On Snapchat, users must be bidirectional friends with each other before they can correspond with each other. Though having a Bitmoji avatar is extremely common on Snapchat, sending a Bitmoji sticker is not. Only 1.6\%, or 3.1 billion, of all Snapchat correspondences consist of Bitmoji stickers, either through embedding a sticker in a photo/video Snap or through directly sending a sticker in the chat box. However, a majority, or 65\%, of users received at least one Bitmoji sticker during the periodo covered by our dataset. For each user, we collect their demographic information including their gender, age, country, and language locale. We also collect their friend count and account age. The correspondence usage statistics are aggregated daily.

\section{RQ1: Reciprocity and Selectivity}\label{sec:rq1}
In this section, we characterize the types of Bitmoji sticker users on Snapchat on dimensions of reciprocity and selectivity. This section is limited to users who use both Snapchat and Bitmoji stickers frequently, since users who don't use Snapchat or Bitmoji stickers frequently do not have enough data points to characterize their usage behavior.

\subsection{Measuring Reciprocity and Selectivity}\label{sec:measure_reciprocity_selectivity}
The exchange of Bitmoji stickers among users can be treated as edges in a weighted, directed network. We want to suitably measure the reciprocity of the Bitmoji stickers a user sends out given how many Bitmoji stickers the user receives from their correspondences. 
We employ the reciprocity metric defined in \citet{squartini2013reciprocity} for weighted networks. For each pair of users $i$ and $j$, let $w_{\vv{ij}}$ be the number of Bitmoji stickers that user $i$ sent to user $j$. Similarly let $w_{\vv{ji}}$ be the number of Bitmoji stickers that user $j$ sent to user $i$. The reciprocated weight is defined as 
\begin{equation}
    w_{\overleftrightarrow{ij}}=\min\{w_{\vv{ij}},w_{\vv{ji}}\}=w_{\overleftrightarrow{ji}}.
\end{equation}
The reciprocity index of user $i$ is therefore 
\begin{equation}
    r_i=\frac{\sum_j w_{\overleftrightarrow{ij}}}{\sum_j w_{\vv{ji}}}, j \text{ is a correspondence of } i.
\end{equation}
This reciprocity index is easily interpretable, ranging from 1 when the user is perfectly reciprocal to 0 when the user displays no reciprocity at all. Since we limit these calculations to active users of Bitmoji stickers, the denominator is always greater than 0 and therefore the reciprocity index is suitably defined.  We note that our definition of reciprocity, as the ratio of the sums, would be more sensitive to how many Bitmoji stickers a user reciprocates in response to all of their received Bitmoji stickers, regardless of which friend they are corresponding with. 

To measure the selectivity $s_i$ of a user $i$, we measure the log-standard deviation of a user’s Bitmoji sticker usage across all their correspondences. Different from the reciprocity index, the selectivity index is calculated from data points belonging to one user only, without needing their correspondences' Bitmoji sticker usage. Example calculations of the reciprocity and selectivity indices can be found in Fig. \ref{fig:reciprocity_selectivity_toy}.
\begin{figure*}
    \centering
    \includegraphics[width=0.85\textwidth]{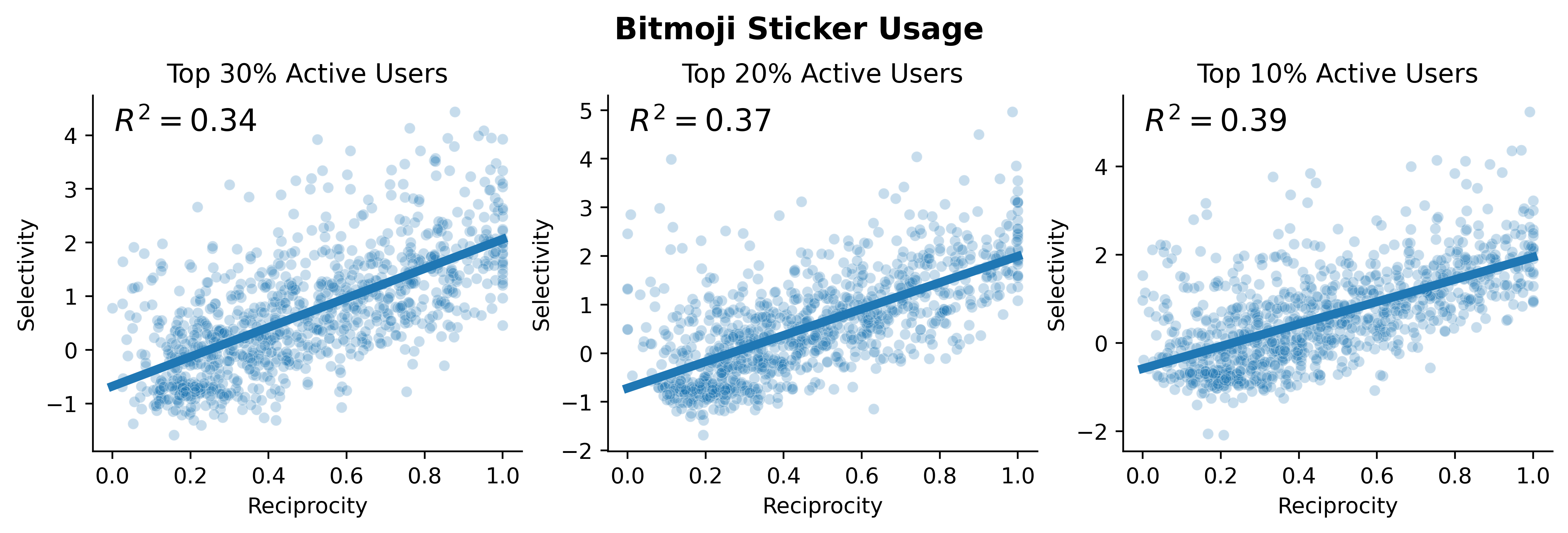}
    \caption{Scatter and regression plots of the Bitmoji sticker reciprocity and selectivity indices for active users of Bitmoji stickers across various levels of overall Snapchat activeness. We find that reciprocity and selectivity indices are positively correlated.}
    \label{fig:reciprocity_selectivity}
\end{figure*}
\subsection{Users on Dimensions of Reciprocity and Selectivity}
We consider users who are active on Snapchat and active in terms of using Bitmoji stickers. To satisfy the frequent Bitmoji sticker use criterion, the user has to have sent more than 15 Bitmoji stickers (above the 90th percentile of all users) in total throughout the month of May. To satisfy the frequent Snapchat use criterion, users included have to be among the top $n\%$ in terms of both the total number of Snapchats sent and the number of correspondences they have. We explore $n=30\%, 20\%, 10\%$ for robustness. The reciprocity and selectivity indices of all eligible users in each setting of Snapchat activeness are displayed in Fig. \ref{fig:reciprocity_selectivity}. In all three settings of Snapchat activeness, there is a positive link between reciprocity and selectivity, with $R^2$ values ranging between 0.34 and 0.39. As a result, it may be informative to place frequent users on a spectrum from being nonreciprocal and unselective (such as Ron in Fig. \ref{fig:reciprocity_selectivity_toy}) to reciprocal and selective (such as Hermione in Fig. \ref{fig:reciprocity_selectivity_toy}).

Now imagine a user who is nonreciprocal but selective. To be nonreciprocal, the user must be sending Bitmoji stickers to their friends at a substantially different level from their friends, either much lower or much higher. To be selective, they must be sending different amounts of Bitmoji stickers to all their friends. As a result, they will be sending \textit{fewer} Bitmoji stickers to people who they receive \textit{more} Bitmoji stickers from, and \textit{more} Bitmoji stickers to people who they receive \textit{fewer} Bitmoji stickers from. This type of people who display the inverse of reciprocity are likely rare \cite{gintis2000strong,fehr2000fairness,smith2014social,eastwick2009reciprocity,eastwick2007selective}.


\section{RQ2: Homophily in Bitmoji Sticker Usage Frequency}
Drawing on prior social network homophily literature and the indication of user-user reciprocity in Bitmoji sticker usage we found from \S\ref{sec:rq1}, we hypothesize that users' Bitmoji sticker usage frequencies could be further related to all of their friends. This section is devoted to examining whether Bitmoji sticker usage frequencies exhibit patterns of network homophily. That is, do friends use Bitmoji stickers at similar frequencies? 

\subsection{Network Assortativity}

Assortativity is a commonly used method to measure network homophily. It is computed as the Pearson correlation of attributes between nodes that share an edge.  For instance, the degree assortativity is calculated as the Pearson correlation coefficient of the degrees of every pair of nodes connected by an edge. The assortativity value ranges between $-1$ and 1 \cite{newman2003mixing}. If the assortativity of a network is greater than 0, then the nodes display preferences to attach to other nodes that are similar in terms of that attribute. If it is less than 0, then the network is disassortative. A common social network assortative attribute is degree assortativity, meaning nodes tend to be preferentially attached to other nodes with similar degree values \cite{newman2002assortative}.

We measure the network assortativity of both (1) the absolute number of Bitmoji stickers sent per user and (2) the proportion of Bitmoji stickers sent out of all Snapchats sent. We also compare this with other attributes that we presume to be assortative, including assortativity of the node degrees \cite{newman2002assortative} and the number of Snapchats sent. We confirm this on both the correspondence network (weighted and undirected) and the friendship network (unweighted and undirected).

\begin{table}[]
    \centering
    \caption{Network assortativity measures indicate that the assortativity of Bitmoji stickers is significantly positive, albeit weaker than other measures of assortativity such as degree or number of Snapchats. All correlations are significant at the 0.001 level.}
    \begin{tabular}{rll}
        \toprule
         & Friendship & Correspondence \\
         & Network  & Network \\
         \midrule
         Degree & 0.340 & 0.122 \\ 
         No. Snapchats & 0.163 & 0.180 \\
         \% Bitmoji Stickers & 0.063 & 0.150\\
         No. Bitmoji Stickers & 0.041 & 0.060 \\
         \bottomrule
    \end{tabular}
    \label{tab:assortativity}
\end{table}

\begin{figure*}
    \centering
    \includegraphics[width=0.8\textwidth]{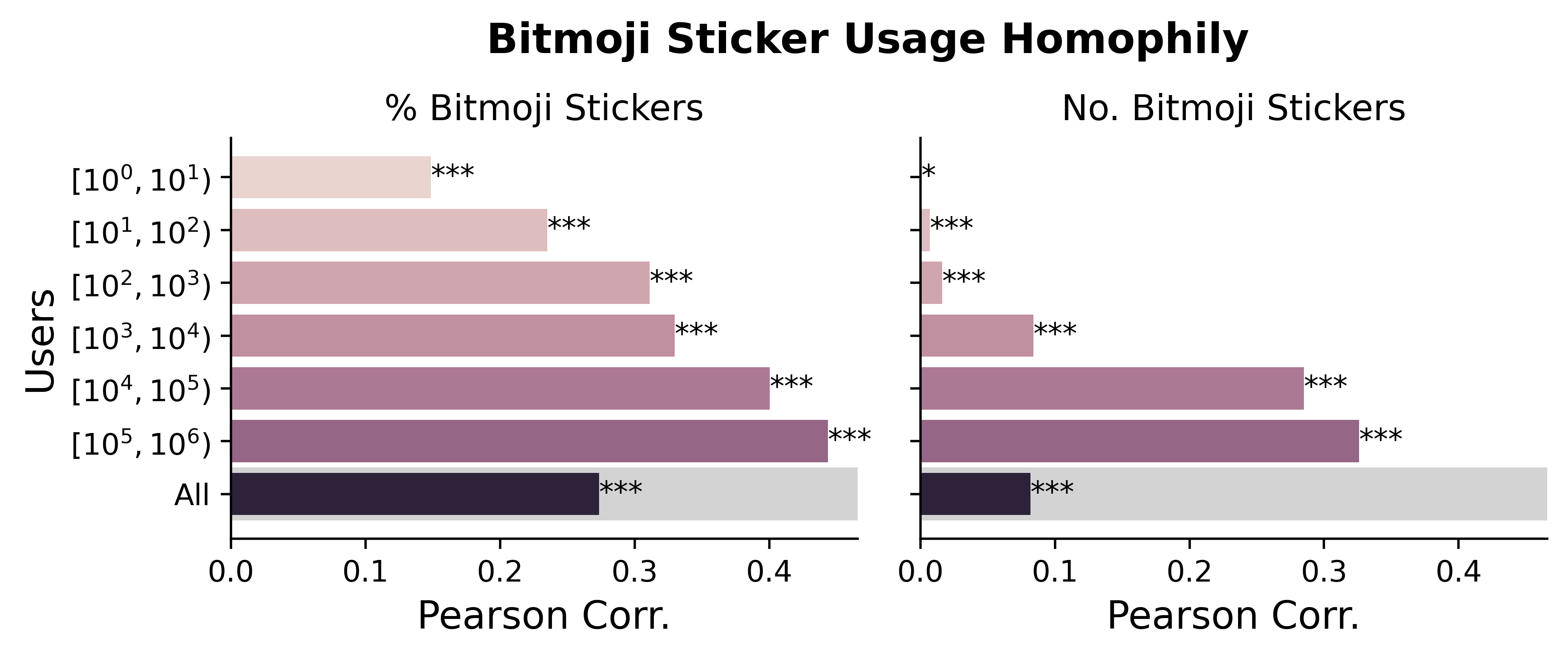}
    \caption{The Pearson correlation of a user's Bitmoji sticker usage (absolute numbers and proportion of total sends) and the average of their friends using the weighted correspondence network. Users are divided into 6 bins depending on total Snapchat usage (number of Snapchats sent). We also compute the overall correlation for all users. $* p<0.05, *** p< 0.001$.}
    \label{fig:homophily}
\end{figure*}
The results (Table \ref{tab:assortativity}) illustrate the assortativity of the various network attributes. We show that node degree is the most assortative attribute, at 0.340 and 0.122 for the friendship and correspondence networks, respectively. This is followed by the assortativity of the number of Snapchats sent. We do indeed show that Bitmoji sticker usage is positively assortative ($p<0.001$), albeit not as assortative as the number of Snapchats sent or node degrees.

\subsection{Comparing a User with Their Friend Group}

Network assortativity treats each edge individually. However, a user can have multiple friends, with varying degrees of interaction. Therefore, we compare how often a user uses Bitmoji stickers with their entire friend group, weighted by the degree of their interactions. Let $b_i$ represent user $i$'s total Bitmoji sticker usage, either as the absolute or proportion of the number of Bitmoji stickers sent.
Let $B_i$ be the weighted average of user $i$’s friends' Bitmoji sticker usage:
\begin{equation}
    B_i = \frac{\sum_j b_ja_{ij}}{\sum_j a_{ij}}, j \text{ is a correspondence of }i,
\end{equation}
where the weight  $a_{ij}$ represents the level of interaction between users $i$ and $j$. In the weighted correspondence network, $a_{ij}$ represents the number of total Snapchats they sent to each other. We then compute the Pearson correlation coefficient between $b_i$ and $B_i$. Since Bitmoji sticker usage could be confounded by total Snapchat usage, we further bucket users by their activity level, disaggregating users who used Snapchat $<10$ times, $<100$ times, $<1000$ times, and so on. 

The results are shown in Fig \ref{fig:homophily}. We observe substantial effects of homophily. For all users, the correlation between their Bitmoji sticker usage and their friends' Bitmoji sticker usage is significantly positive at 0.27 ($p<0.001$) for the proportion of Bitmoji stickers used and 0.18 for the absolute number of Bitmoji stickers used ($p<0.001$). We also find that the correlations are statistically significant for users of all activeness, with the degree of homophily increasing monotonically as the user activeness grows, reaching 0.44 ($p<0.001$) for the proportion of Bitmoji stickers used and 0.33 ($p<0.001$) for the absolute number of Bitmoji stickers used. These analyses were further replicated with the unweighted version of the correspondence network and the friendship network, in which cases $a_{ij}=1$ for all linked user pairs, with similar results.

\begin{figure}
    \centering
    \begin{minipage}{0.5\textwidth}
         \includegraphics[width=\textwidth]{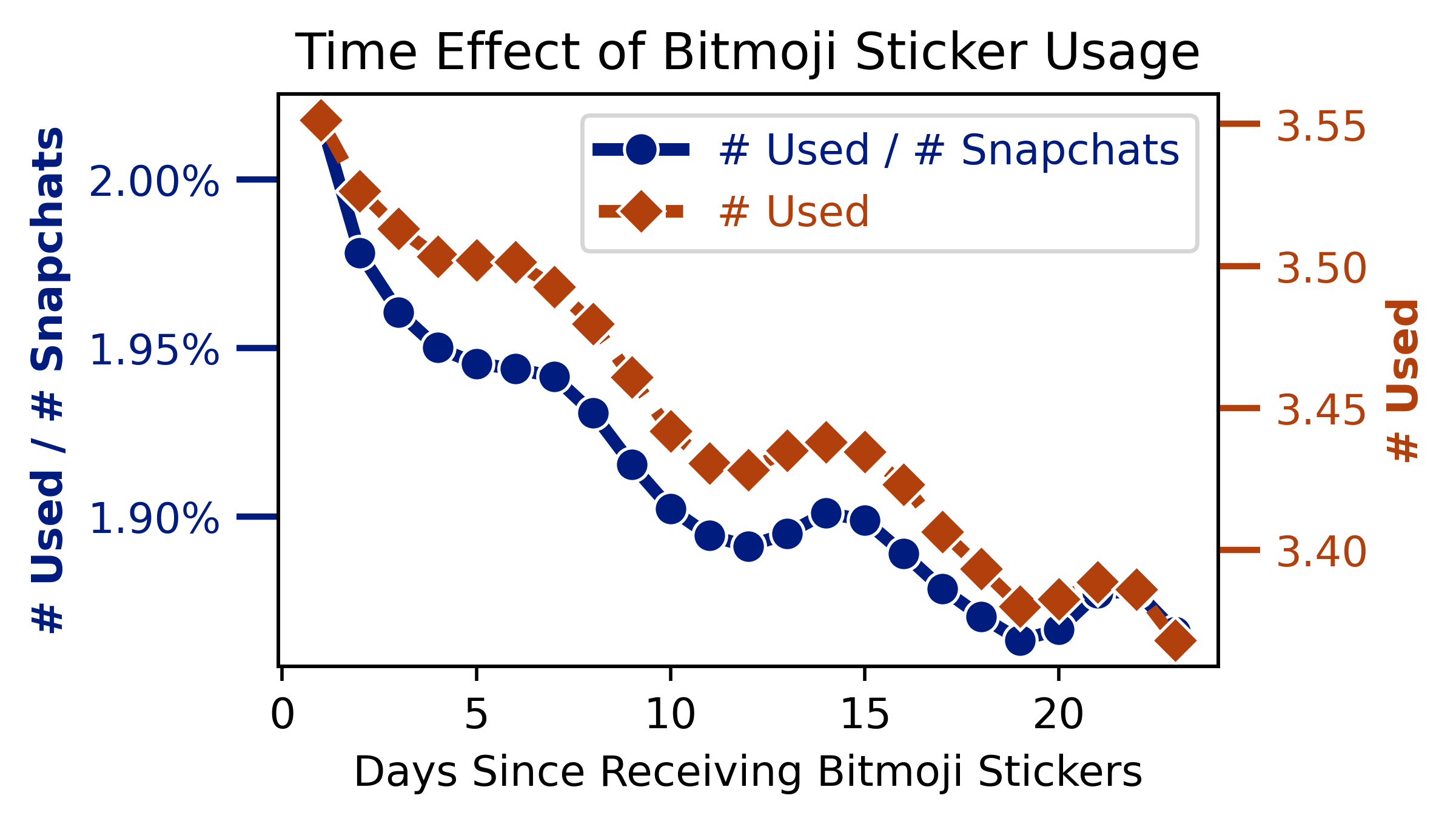}
         \caption{The average normalized (left axis) and absolute (right axis) number of Bitmoji stickers given the number of days after receiving a Bitmoji sticker. Please note the y-axis range: Bitmoji stickers are used in 1.6\% of all messages on average.}
         \label{fig:time_effect}
     \end{minipage}
     \hfill
     \begin{minipage}{0.45\textwidth}
        \includegraphics[width=\textwidth]{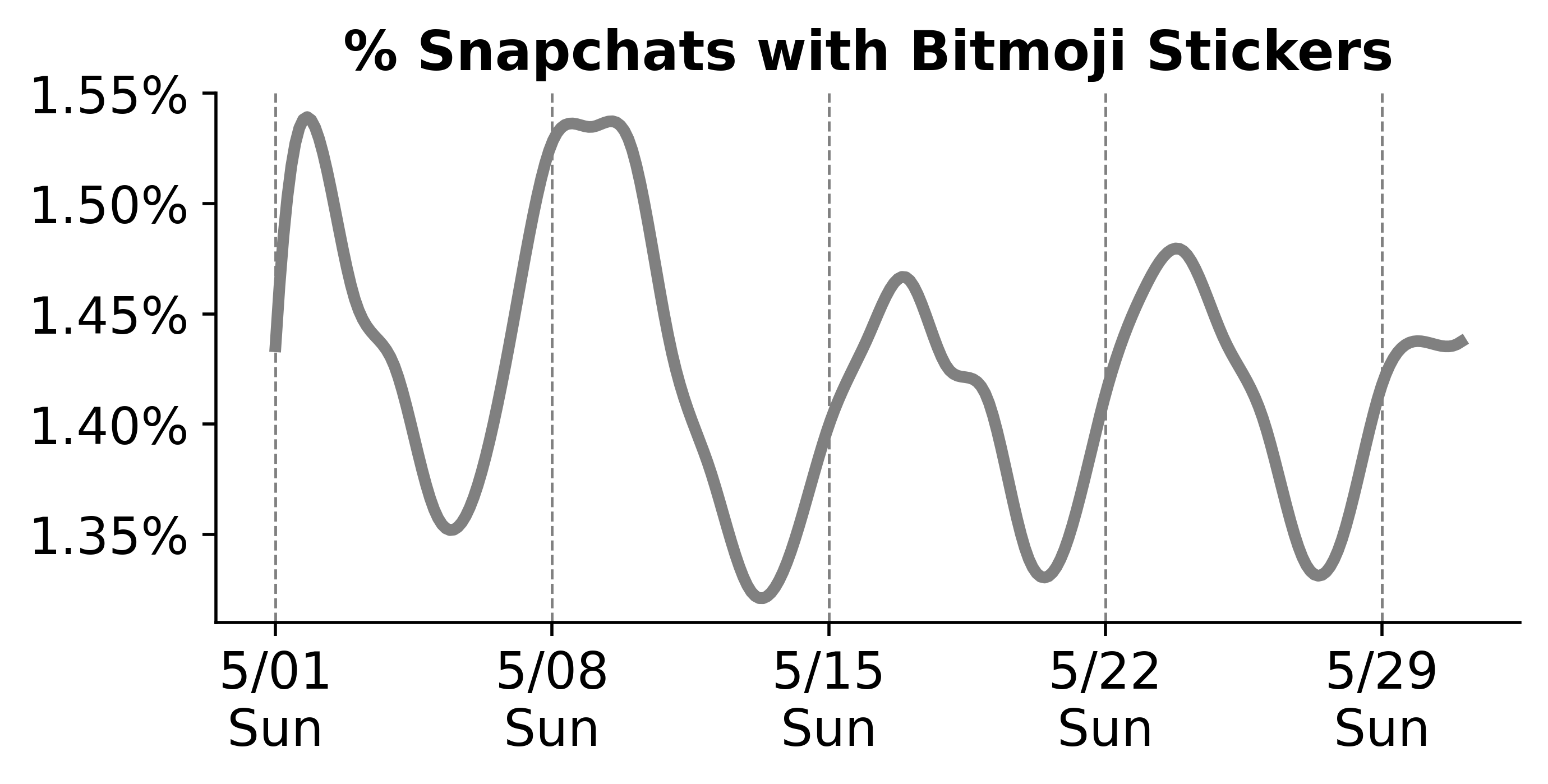}
    \caption{The proportion of Snapchats that have Bitmoji stickers over time displays patterns of weekly cyclicity. Please note the y-axis range.}
    \label{fig:weekly}
    \end{minipage}
\end{figure}
\section{RQ3: Effects of Receiving Bitmoji Stickers}
Knowing that Bitmoji sticker usage is homophilous, we want to explore to what extent this could be induced by social contagion effects, in which seeing a friend use Bitmoji stickers could motivate that user to also use Bitmoji stickers. Consequently, we measure how receiving Bitmoji stickers impacts a user's future in-app behavior, both for using Bitmoji stickers and for using Snapchat overall. We use a quasi-experimental method, in which users are considered `treated’ if they received Bitmoji stickers from their friends during a treatment period. We answer two experimental questions:
\begin{itemize}
    \item \textbf{\texttt{E1 Bitmoji}:} Does receiving Bitmoji stickers drive future Bitmoji sticker usage for users who continue to use Snapchat?
    \item \textbf{\texttt{E2 Snapchat}:} Does receiving Bitmoji stickers drive overall in-app engagement and user retention on Snapchat?
\end{itemize}

\subsection{Quantifying Treatment Effect} To examine the effects of receiving Bitmoji stickers on future Bitmoji sticker usage and overall in-app behavior, we define and compute outcomes metrics that quantify usage frequencies. For \texttt{E1 Bitmoji}, we compute three measures of Bitmoji usage: the absolute number of Bitmoji stickers used, the proportion of Snapchats containing Bitmoji stickers, and the proportion of users who used Bitmoji stickers for the first time in the month of May. Note that we consider a user’s Bitmoji sticker usage with all of their friends, not only the ones who sent them Bitmoji stickers. For \texttt{E2 Snapchat}, we compute two measures of Snapchat usage: the number of Snapchats sent and the proportion of users who used Snapchat (users retained).

\subsection{Preliminary Results}
Fig. \ref{fig:time_effect} shows the number of Bitmoji stickers sent to any correspondent as a function of the number of days that have passed since receiving a Bitmoji sticker from a correspondent on day 0. The blue line shows the number of Bitmoji stickers used, normalized by the total number of Snapchats sent, and the orange line shows the absolute number of Bitmoji stickers used. Both lines demonstrate that Bitmoji sticker usage declines with time since receiving a Bitmoji sticker, which could suggest that users are less likely to use Bitmoji stickers as time progresses from the day they received a sticker. Moreover, we observe slight fluctuations in the Bitmoji sticker usage that could be explained by weekly cyclicity (see Fig. \ref{fig:weekly}). These preliminary findings motivate us to further investigate whether receiving Bitmoji stickers from a friend can impact future user behavior.

\subsection{Propensity Score Matching}
To more rigorously examine how receiving a Bitmoji sticker influences future Bitmoji sticker (\texttt{E1}) and Snapchat (\texttt{E2}) usage, we adopt a quasi-experimental approach for observational data using propensity score matching \cite{rosenbaum1984reducing,gelman2007causal,imbens2015causal}. Within the constraints of observational data, we can use the propensity score matching technique to match similar groups of \texttt{Treated} and \texttt{Control} users together to estimate the effects of the treatment. The intuition is to algorithmically predict the propensity for a user to receive the treatment given some prior covariates. Thus, \texttt{Treated} and \texttt{Control} users with similar propensity scores will have similar covariates, enabling us to compare their treatment outcomes. Any conclusions we draw about the causal effects, as a result, are most likely not due to the covariates we account for. 
\begin{figure*}
    \centering
    \includegraphics[width=\textwidth]{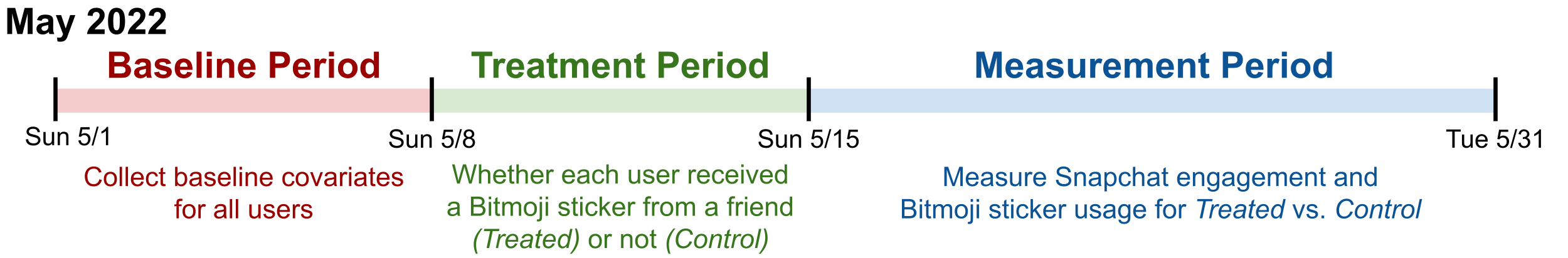}
    \caption{Illustration of the workflow used to examine the effects of receiving Bitmoji stickers from friends (treatment threshold $\theta=1$ shown).}
    \label{fig:psm_illustration}
\end{figure*}
\subsubsection{Experimental Setup} For our purposes, we split the data into three time periods. See Fig \ref{fig:psm_illustration} for a schematic illustration.  The baseline period is the first week of May. We use data from this baseline period to collect Snapchat usage features that could be relevant to how likely users are to receive the treatment later (e.g., the number of Snapchats sent, the number of Bitmoji stickers sent/received). The treatment period is the second week of May. During this period, we consider a user as \texttt{Treated} if they received at least $\theta$ Bitmoji sticker(s) from a friend and \texttt{Control} if they didn't. In our experiments, we fixed $\theta=0$, but we also tested $\theta=1, 2$ for robustness. Finally, we designate the rest of May as the measurement period. We match similar users from the \texttt{Treated} and \texttt{Control} groups to estimate the effect of treatment, which is receiving a Bitmoji sticker. We use weeks as the resolution of time periods for two reasons. First, since Bitmoji sticker usage is low at below 2\% of all Snapchats, we want to expand the treatment period to allow for more users to be deemed \texttt{Treated}, increasing our sample size. Second, there is strong evidence of weekly cyclicity in normalized Bitmoji sticker usage (Fig. \ref{fig:weekly}). By considering resolutions at the weekly level, we can avoid confounders induced by weekly behavioral patterns.

The number of total users included in the \texttt{E1 Bitmoji} and \texttt{E2 Snapchat} experiments is 177 million and 181 million, respectively. 42\% of the \texttt{E1 Bitmoji} users and 41\% of the \texttt{E2 Snapchat} users are \texttt{Treated}. To be included in the \texttt{E1 Bitmoji} experiments, the user must have used Snapchat at least once during each of the three periods. The constraint is slightly relaxed for the \texttt{E2 Snapchat} experiments in that the user does not have to have used Snapchat during the measurement period. This is because we want to measure how many Bitmoji stickers will be used by an \textit{active} user of Snapchat in \texttt{E1 Bitmoji}, but we want to measure the overall in-app engagement and user retention in \texttt{E2 Snapchat} including those who no longer used Snapchat. In either type of experiment, a user is \texttt{Treated} if they received at least $\theta$ Bitmoji stickers during the treatment period.

\begin{figure*}
    \centering
    \includegraphics[width=0.7\textwidth]{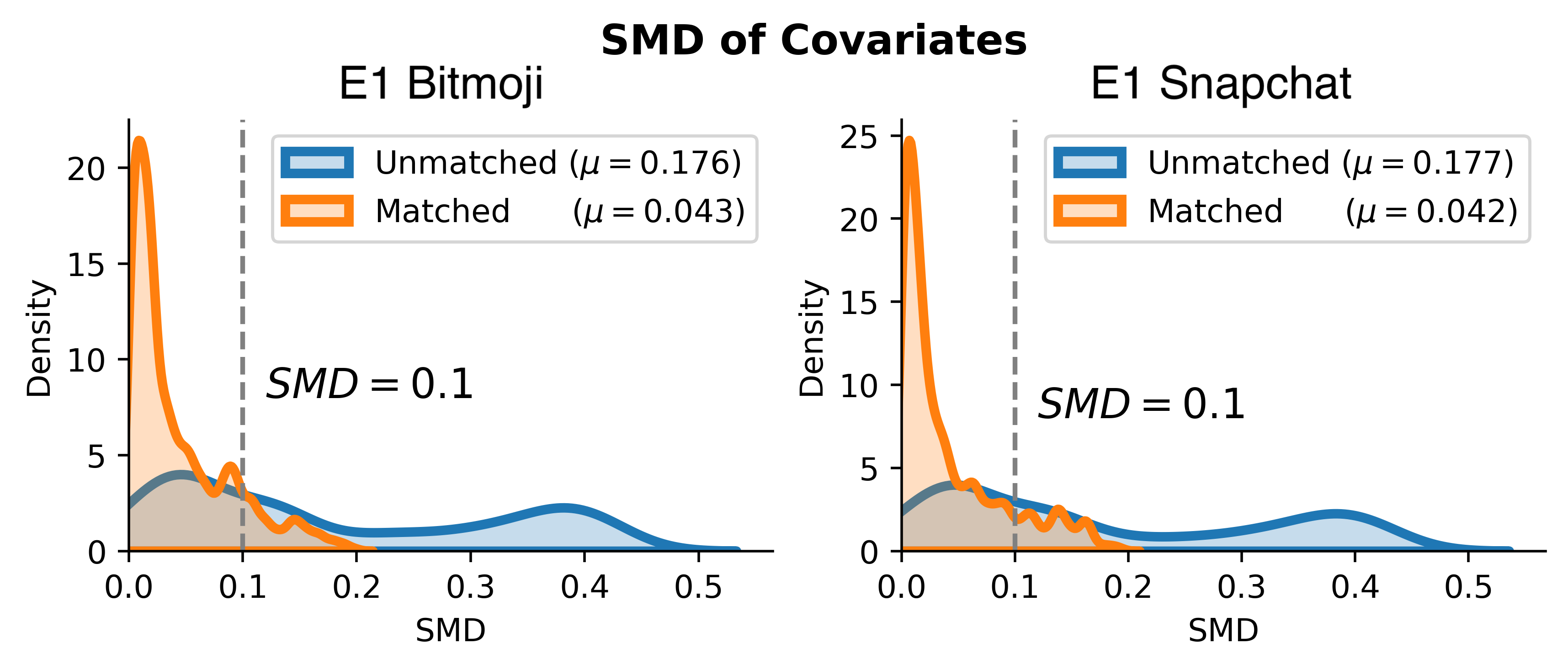}
    \caption{Standardized mean differences (SMD) for covariates before and after matching for \texttt{E1 Bitmoji} (left) and \texttt{E2 Snapchat} (right). Matching substantially reduces biases in the covariates.}
    \label{fig:smd}
\end{figure*}
\subsubsection{Covariates.} The covariates should be carefully chosen so that differences in confounding variables between the matched groups of \texttt{Treated} and \texttt{Control} users are minimized. We use four sets of covariates. The first set of covariates includes the users' information, including their age, gender, country, language locale, Snapchat account age, and the number of friends on Snapchat. Country and language locale features are one-hot encoded. The second set of covariates consists of Snapchat usage statistics during the baseline period, including the total number of Snapchats that are sent or received (separately for photo/video Snaps or text-based chats), how many of those were group correspondences versus one-on-one correspondences, how many included Bitmoji stickers or other types of stickers, and how many correspondences they had. The third set of covariates includes the reciprocity and selectivity indices of Bitmoji sticker usage during the baseline period, as described in \S\ref{sec:measure_reciprocity_selectivity}. They are set to $-1$ if they are undefined. The reciprocity index is undefined when the user sent no Bitmoji stickers (the denominator would be zero); the selectivity index, which is the standard deviation, is undefined when the user has only one correspondence.  The last set of covariates includes the number of Snapchats they sent and received during the treatment period since that could considerably influence their propensity to receive the Bitmoji sticker treatment. 
\

\subsubsection{Propensity Score Modeling.} Traditionally, propensity score models are logistic regression models \cite{stuart2010matching,austin2011introduction}. In this work, we also explored nonlinear models in the form of deep neural networks (DNNs) \cite{setoguchi2008evaluating}. The DNN we use consists of 5 hidden layers (dimensions: 256, 128, 64, 32, 16), all with ReLu activation, followed by a single output layer with sigmoid activation suitable for binary classification outputs \cite{sharma2017activation}. The weights are optimized using an Adam optimizer (learning rate = 0.001) \cite{kingma2014adam} with binary cross-entropy as the loss function. Due to the large data size, a fixed 10\% of the data was sampled for training in each experimental setup. Both the logistic regression and DNN models were trained for at most 10 epochs with early stopping mechanisms that monitor the training losses. In reality, all models completed within 5 epochs. To determine model fit, the models are evaluated with accuracy, AUC, and macro-F1 on the training set. Unlike a traditional machine learning model, we are not interested in the inference capabilities of the model but in how well the model can balance propensities, which is measurable via training dataset fit. For both $\theta = 0$ and $\theta  = 1$, we find that the DNN model outperforms the logistic regression model in terms of both model fit and covariate balancing of matched samples. Therefore, we chose DNN as our model. We apply the trained DNN model to all users in our experiments to obtain propensity scores for each user.

\subsubsection{Matching}
Using the estimated propensity scores, we can match users from the \texttt{Treated} group with users from the \texttt{Control} group with similar propensity scores together. To take full advantage of our large dataset and to avoid problems and computational overhead of exact matching, we use subclassification \cite{stuart2010matching,olteanu2017distilling,king2019propensity,saha2021advertiming,bisberg2022gift}. We first check whether any users have propensity scores that are two standard deviations away from the mean \cite{saha2019social}. There are none in any of our experimental setups. We then divide users into 100 strata of equal widths according to their propensity scores \cite{saha2021advertiming,bisberg2022gift}. To ensure the validity between matched groups, we remove strata with insufficient users from either the \texttt{Treated} or \texttt{Control} group  \cite{saha2021advertiming,bisberg2022gift}. We measure the quality of our matching by computing the standardized mean differences (SMD), also known as Cohen’s \textit{d}, for each
covariate, which is the difference in the mean covariate of the two groups divided by their pooled standard deviation \cite{stuart2010matching,austin2011introduction}. The lower the SMD, the better the balance. In our experiments, we require all covariates to have SMDs less than 0.2, because two groups are considered balanced if the SMDs of all their covariates are less than 0.2 because the differences between them are negligible \cite{stuart2010matching,austin2011introduction}. We further require a minimum sample size for each \texttt{Treated} or \texttt{Control} group in each stratum to be 10,000 to ensure having sufficient data points for comparison. Strata that violate either of these requirements are removed from our analysis. In a later section, we will test the robustness of our models with varying levels of strata validity requirements.

These strata validity requirements are satisfied by most of the strata. There are 70 valid strata for \texttt{E1 Bitmoji} with 52 million \texttt{Treated} users and 73 million \texttt{Control} users. For \texttt{E2 Snapchat}, we have 75 valid strata with 107 million \texttt{Treated} and 166 million \texttt{Control} users. The SMDs of covariates before and after matching are shown in Fig. \ref{fig:smd}. There is a 75.38\% and 76.49\% reduction in SMDs for E1 and E2, respectively, resulting in SMDs of matched samples averaging 0.043 (E1) and 0.042 (E2). We thus deem the matched samples of \texttt{Treated} and \texttt{Control} users balanced in terms of covariates and continue our analysis.
\begin{figure*}
    \centering
    \includegraphics[width=0.7\textwidth]{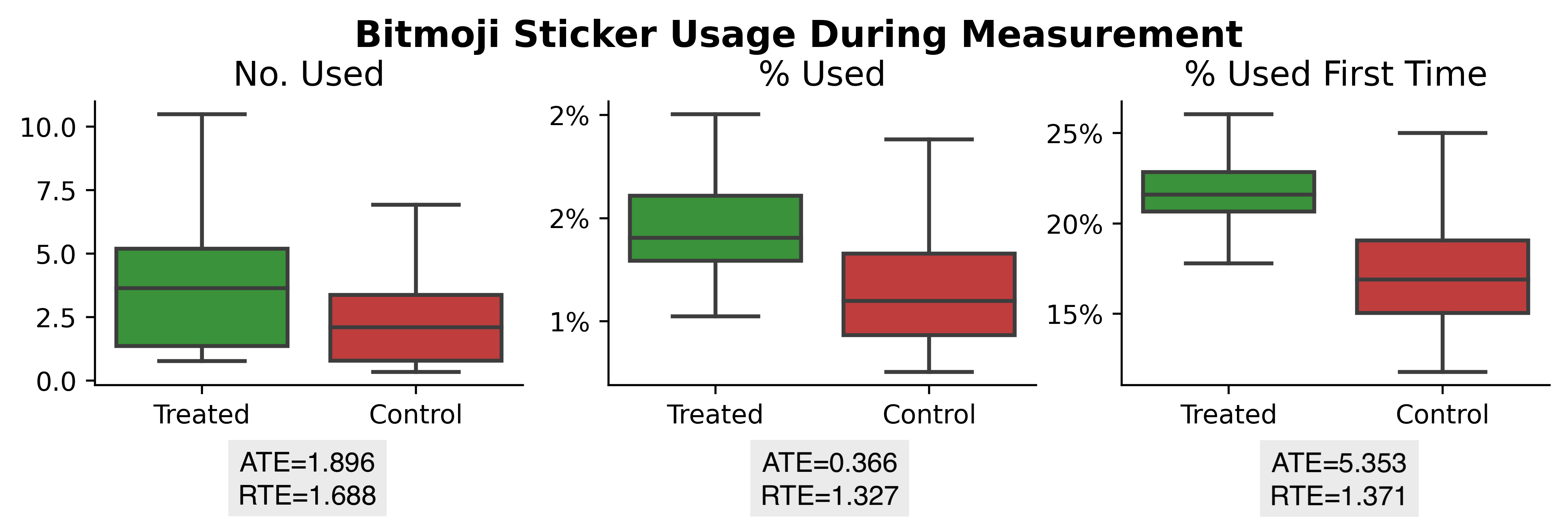}
    \caption{Treatment effect estimation for \texttt{E1 Bitmoji}. Receiving a Bitmoji sticker during the treatment period leads to higher Bitmoji sticker usage during the measurement period.  }
    \label{fig:e1}
\end{figure*}
\subsection{Impact on Future Snapchat Engagement}
\begin{figure}
    \centering
    \includegraphics[width=\linewidth]{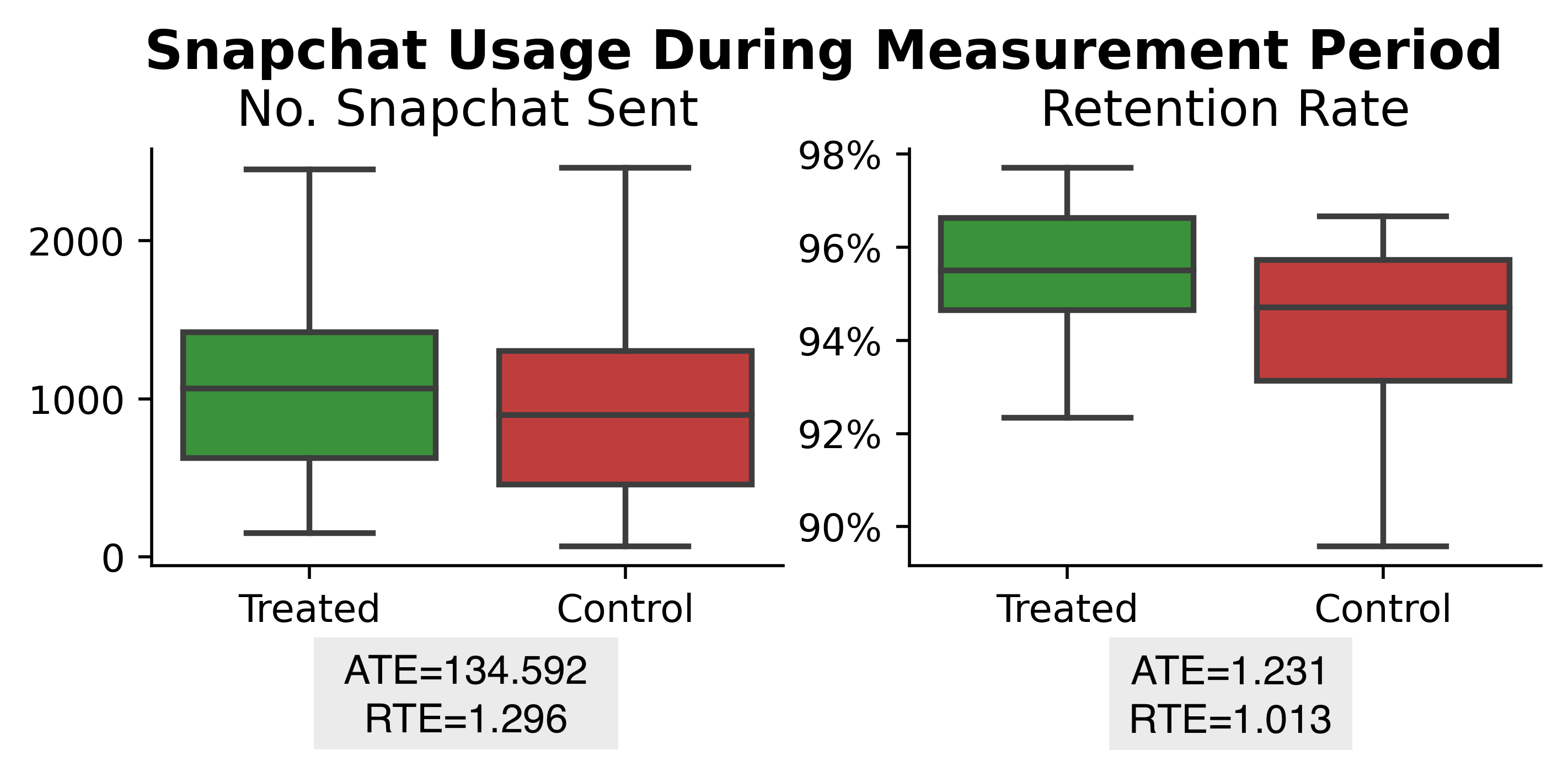}
    \caption{Treatment effect estimation for \texttt{E2 Snapchat}. Receiving a Bitmoji sticker during the treatment period leads to higher overall in-app engagement during the measurement period. }
    \label{fig:e2}
\end{figure}
To estimate the treatment effect, we compute the average treatment effect (ATE), which is the difference in the means of the outcomes between the \texttt{Treated} and the \texttt{Control} groups, and the relative treatment effect (RTE), which is the ratio of the means of the outcomes of the \texttt{Treated} group to the means of the outcomes of the \texttt{Control} group. Both the ATE and RTE are calculated per stratum and then averaged, weighted by the number of users per stratum. An RTE greater than 1 or ATE greater than 0 indicates that the \texttt{Treated} group had a higher treatment outcome than the \texttt{Control} group, and vice versa for an RTE $< 1$ or ATE $< 0$. We conducted paired $t$-tests to determine the significance of the difference between outcome means and Kolmogorov-Smirnov (KS) tests to measure the likelihood that the outcomes of \texttt{Treated} and \texttt{Control} are from the same distribution over the strata.

\subsubsection{E1 Bitmoji Results} Fig. \ref{fig:e1} displays the results for \texttt{E1 Bitmoji}. Across all three measures of Bitmoji sticker usage, the RTEs are $>1$ and the ATEs are positive, indicating receiving a Bitmoji sticker during the treatment period (i.e., treated) resulted in higher Bitmoji sticker use in the measurement period. The differences are statistically significant for the absolute number of Bitmoji stickers used ($t$-test  $t=8.217$, $p<0.001$; KS-test $D=0.314$, $p<.01$) and in the proportion of Snapchats containing Bitmoji stickers ($t$-test  $t=15.092$, $p<0.001$; KS-test  $D=0.614$ $p<.001$). We also see that an average of 25.2\% of the \texttt{Treated} users used Bitmoji stickers for the first time during the measurement, which is significantly higher than the 14.8\% of the users from the \texttt{Control} group ($t$-test $t=17.537$,  $p<0.001$; KS-test $D=0.600$, $p<.001$).

\subsubsection{E2 Snapchat Results} In terms of overall in-app engagement (\texttt{E2 Snapchat}), we observe similar differences between the \texttt{Treated} and \texttt{Control} user groups in Fig. \ref{fig:e2}. Users who received Bitmoji stickers in the treatment period sent more Snapchats (RTE > 1) later on. However, while this difference is significant using the $t$-test ($t=9.848$, $p<0.001$), it is not significant using the KS-test ($D=0.173$). We do however find that 95.4\% of the \texttt{Treated} continued using the app during the measurement period, compared to 94.2\% of the \texttt{Control} group, a small but statistically significant difference ($t$-test $t=6.162$, $p<0.001$; KS-test $D=0.280$, $p<0.01$). The results mostly support our hypothesis that receiving Bitmoji stickers will lead to higher overall in-app engagement.

\bigskip
\noindent
There could be many explanations for these results. First and foremost, users may have exerted social influence on their friends by sending Bitmoji stickers, motivating their friends to use more Bitmoji stickers themselves in the future, and, by extension, Snapchat overall. Users also may not have been aware of how quirky, funny, or otherwise interesting the Bitmoji stickers are until they saw their friends use them.
Though we made effort to reduce observable covariates that could be explained by homophilous preferences for using Bitmoji stickers, there could be latent homophily that contributes to users' propensity to use Bitmoji stickers \textit{and} also their likelihood to become friends \cite{shalizi2011homophily}. Finally, some treated users could be using Bitmoji stickers with their friends as a form of reciprocation. To this end, we conduct further analyses into how behavioral reciprocity and selectivity impact users’ future behavior.
\begin{table*}[]
    \centering
    \caption{The effects of receiving a Bitmoji sticker during the treatment period on the number of Bitmoji stickers used (\texttt{E1 Bitmoji}) and the number of Snapchats sent (\texttt{E2 Snapchat}) during the measurement period are different for each type of Bitmoji sticker user.}
    
    \label{tab:user_types_e1_e2}
    \begin{tabular}{llrrrrrllll}
        \toprule
         & User Type &   \# \texttt{Treated} &  \# \texttt{Control}& \texttt{Treated} $\mu$ & \texttt{Control} $\mu$ & ATE & RTE & $t$-test& KS test\\
         \midrule 
         \midrule
         \multicolumn{3}{l}{\texttt{\textbf{E1 Bitmoji}}}  \\
         & Infrequent & 47M& 72M & 3.649 & 1.572 & 2.077& 2.462  \BarGrey{4.92} & 11.026*** & 0.722*** \\
         & $\neg(R\&S)$ & 4.9M & 1.5M & 25.262 & 16.945 & 8.316 & 1.487  \BarGrey{2.97} & 24.477*** & 0.870***\\
         & $R\&S$ &2.6M&540K&  33.665 & 16.180 & 17.484 & 2.049  \BarGrey{4.10} & 9.151*** & 0.789**\\
         \midrule
         \multicolumn{3}{l}{\texttt{\textbf{E2 Snapchat}}} \\
         & Infrequent &  93M& 164M& 1100.515 & 892.904 & 207.611 & 1.532 \BarGrey{3.06} & 3.091** & 0.360*** \\
         & $\neg(R\&S)$ & 8.5M& 3.6M&  1981.217 & 1672.921 & 308.296 & 1.248 \BarGrey{2.50} & 6.122*** & 0.379*** \\
         & $R\&S$ & 4.4M & 1.1M & 1324.997 & 882.920 & 442.077 & 1.531 \BarGrey{3.06} & 12.336*** & 0.571**\\
         \bottomrule
    \end{tabular}
\end{table*}
\subsection{Usage Impact Based on Bitmoji Sticker User Types} 
To better understand how the process of Bitmoji sticker behavior is spread socially, we want to contextualize how a user's reciprocity tendencies play a role in this process. A reciprocal user might be more influenced by receiving a Bitmoji sticker because they want to reciprocate the behavior. From \S\ref{sec:rq1}, we know that frequent Bitmoji sticker users range from being reciprocal and selective to nonreciprocal and unselective. As such, we can separate the users into three mutually exclusive Bitmoji sticker user types. The first type of users is infrequent users of Bitmoji stickers, defined as users who used less than or equal to 4 Bitmoji stickers (90\% percentile) during the baseline period. The second type of users is frequent Bitmoji sticker users who are also nonreciprocal and unselective users ($\neg(R\&S)$), defined as users who used more than 4 Bitmoji stickers and have reciprocity and selectivity indices both below the medians, which are 0.273 and 0.496, respectively. The third type of users is frequent Bitmoji sticker users who are also reciprocal and selective ($R\&S$), defined as users with reciprocity and selectivity indices larger than or equal to the medians. Users who do not belong to any of the three types are left out of the analysis. We then compare \texttt{Treated} and \texttt{Control} users who are of the same Bitmoji sticker user type and strata. Strata with less than 10,000 users per treatment group and user type are discarded. For each user type in each treatment group, we analyze how many Bitmoji stickers (\texttt{E1 Bitmoji}) and Snapchats (\texttt{E2 Snapchat}) they sent during the measurement period. 

\subsubsection{Impact on Bitmoji Sticker Usage (\texttt{E1})} 
While Table \ref{tab:user_types_e1_e2} shows that receiving Bitmoji stickers significantly increases the number of future Bitmoji stickers used by all types of users, the effects are noticeably different across user types. First, compared to the more frequent users of Bitmoji stickers, infrequent Bitmoji sticker users continue to use Bitmoji stickers infrequently, regardless of their treatment status. That being said, there is a small and significant rise in the number of Bitmoji stickers used, indicating that even users who were not predisposed to use Bitmoji stickers use them more often after receiving them. As for frequent Bitmoji users, the treatment effect on $R\&S$ users dwarfs those of the $\neg(R\&S)$ users with twice as much ATE. These results suggest that users who are reciprocal and selective ($R\&S$) are particularly impacted by receiving Bitmoji stickers and will use substantially more Bitmoji stickers in the future. It is possible, being they are reciprocal, that they used more Bitmoji stickers with the same friends who sent them Bitmoji stickers.

\subsubsection{Impact on Snapchat Engagement (\texttt{E2})}
Looking at how receiving a Bitmoji sticker impacts overall in-app engagement for each user type reveals additional insights (Table \ref{tab:user_types_e1_e2}). All user types experience higher Snapchat engagement after receiving a Bitmoji sticker, including the infrequent Bitmoji sticker users (ATE$=207.611$, RTE$=1.532$). The treatment effect is larger for $R\&S$ users (ATE$=442.077$, RTE$=1.531$) than it is for $\neg(R\&S)$ users (ATE$=308.296$, RTE$=1.248$). All outcome differences are statistically significant at the 0.01 level using both paired $t$-tests and KS tests. As a result, we argue that compared to nonreciprocal and unselective users ($\neg(R\&S)$), reciprocal and selective users ($R\&S$) are more sensitive to receiving Bitmoji stickers, responding with higher Snapchat usage.

\begin{table*}[]
    \centering
    
    \caption{Robustness analysis examining the treatment effects for varying treatment thresholds $\theta$ (*** $p<0.001$, N.S.$=$ not significant). To be considered treated, the user must have received more than $\theta$ Bitmoji stickers during the treatment period.}
    \label{tab:robust_sigma}
    \begin{tabular}{cccccccccc}
    \toprule
    & $\theta$ &  No. Strata & No. \texttt{Treated} &  \texttt{No. Control} & Matched SMD & SMD Reduction &    RTE & $t$-test & KS test\\
    \midrule
    \midrule
    \multicolumn{3}{l}{\texttt{\textbf{E1 Bitmoji}}} \\
    & 1 &        55 &  31,455,336 &   66,864,895 &  0.045 & -74.24\% &  1.912 & *** & *** \\
    & 2 &        31 &  27,705,617 &   27,785,441 &  0.017 & -90.53\% &  1.653 & *** & *** \\
    & 3 &        38 &  14,004,951 &   53,568,098 &  0.026 &  -85.48\% &  1.896 & *** & *** \\
    \midrule
    \multicolumn{3}{l}{\texttt{\textbf{E2 Snapchat}}} \\
    & 1 &        65 &  62,322,200 &  175,487,794 &  0.034 & -80.59\% &  1.153 & *** & *** \\
    & 2 &        31 &  55,419,860 &   57,961,164 &  0.017 & -90.13\% &  1.044 & *** & *** \\
    & 3 &        36 &  30,721,576 &  100,992,300 &  0.025 & -85.82\% &  0.992 & N.S. & N.S.\\
    \bottomrule
    \end{tabular}
\end{table*}

\begin{table}[]
    \centering
    \caption{Strata validity requirements robustness analysis: (a) the minimum number of users required per treatment group in each stratum and (b) the maximum SMD per covariate in each stratum. The treatment threshold $\theta$ is 0. }
    \label{tab:robust}
    \begin{subtable}{0.46\textwidth}
    \caption{Minimum number of users (maximum SMD $=$ 0.2)}
    \vspace{-2mm}
    \label{tab:robust_user}
    \centering
     \begin{tabular}{lllll}
    \toprule
     &  \multicolumn{2}{c}{\texttt{\textbf{E1 Bitmoji}}} & \multicolumn{2}{c}{\texttt{\textbf{E2 Snapchat}}}\\
     \cmidrule(lr){2-3} \cmidrule(lr){4-5}
    Min Users &   No. Strata &        RTE &    No. Strata &        RTE   \\
    \midrule
       1,000 &  72 &  1.692 &  75 &   1.297 \\
      50,000 &  65 &  1.683 &  70 &  1.274 \\
     100,000 &  60 &  1.681 &  65 &  1.255 \\
     200,000 &  46 &  1.669 &  51 &  1.223 \\
    \bottomrule
    \end{tabular}
    \end{subtable}
    \hfill
    \begin{subtable}{0.46\textwidth}
    \caption{Maximum SMD (minimum number of users $=$ 10,000)}
    \vspace{-2mm}
    \label{tab:robust_smd}
    \centering
    \begin{tabular}{lllll}
    \toprule
    &  \multicolumn{2}{c}{\texttt{\textbf{E1 Bitmoji}}} & \multicolumn{2}{c}{\texttt{\textbf{E2 Snapchat}}}\\
     \cmidrule(lr){2-3} \cmidrule(lr){4-5}
    Max SMD &   No. Strata &        RTE &    No. Strata &        RTE   \\
    \midrule
     0.25 &  82 &   1.700 &  83 &  1.332 \\
     0.15 &  39 &  1.628 &  37 &  1.183 \\
      0.10 &  25 &  1.566 &  19 &  1.046 \\
     0.05 &   9 &  1.551 &   4 &  1.039 \\
     \bottomrule
    \end{tabular}
    \end{subtable}
    \vspace{-5mm}
\end{table}

\subsection{Robustness Analysis}
To demonstrate the reliability of our findings, we conduct several robustness checks. In our experiments we set the treatment threshold $\theta$ to be 0, meaning that a user is in the \texttt{Treated} group if they received at least one Bitmoji sticker during the treatment period. Since a user can receive more Bitmoji stickers, we question whether the treatment effects are still observable with higher thresholds. We test with thresholds up to $\theta=3$. Since Bitmoji usage is generally low, testing up to $\theta=3$ covers the majority of the cases. We record these results in Table \ref{tab:robust_sigma}, demonstrating that the RTEs are greater than 1 in almost all experiment settings. The sole exception is the case of $\theta=3$ in \texttt{E2 Snapchat} where the RTE is less than 1, but the differences between the \texttt{Treated} and \texttt{Control} groups are not statistically significant to refute our claims. We also test varying strata validity requirements in Table \ref{tab:robust} by changing the minimum number of users required per treatment group and the maximum SMD per covariate in each stratum. The RTEs are greater than 1 in all cases, and we also verify that the differences are statistically significant at the 0.001 level using both $t$-tests and KS tests. Therefore, we are confident that our results are reliable.

\section{Discussion}

In this work, we consider the social process involved in a style of communication--using Bitmoji stickers with friends. We tackle this problem from three perspectives: reciprocity/selectivity, network homophily, and social contagion. Below we discuss the implications and limitations of our findings.
\subsection{Implications}

\subsubsection{User Reciprocity and Selectivity} Our work reveals that sending Bitmoji stickers ranges from being reciprocal and selective to being nonreciprocal and unselective. While these two measures quantify distinct behavioral characteristics, they are shown to be correlated in this empirical work. Depending on their reciprocity and selectivity tendencies, people are differently impacted by receiving Bitmoji stickers from their friends. Reciprocal and selective users are particularly receptive to the social stimuli, responding with palpably higher future Bitmoji sticker usage and Snapchat usage. Since different users respond differently to social signals, identifying user types can inform platform designers and analysts to cater their designs and solutions to different user personas.

\subsubsection{Behavioral Homophily and Contagion} 
In this paper, we show evidence of network homophily and social contagion. Users use Bitmoji stickers at similar frequencies as their friends and are more likely to use more Bitmoji stickers upon receiving one from their friends. The social contagion theory could explain why we observe the patterns of homophily as users mimic their friends’ behaviors. Social contagion has a largely unconscious influence on human behavior and emotions \cite{marsden1998memetics,christakis2013social}. A rich body of literature has examined the contagious nature of human emotions \cite{barsade2002ripple,ferrara2015measuring,bastiampillai2013depression,rosenquist2011social} as well as behaviors such as smoking \cite{christakis2008collective}, diet \cite{pachucki2011social}, curiosity \cite{dubey2021curiosity}, and generosity  \cite{tsvetkova2014social,pressman2015pif,bisberg2022gift}. Adding on to prior research, our work carries important implications regarding possible evidence of mass behavioral contagion of pictorial communication styles online. It could be argued that some apparent instances of contagion could be due to social conformity or peer pressure \cite{wheeler1966toward}. However, this argument stands only when the behavior in question is categorically negative, such as substance abuse \cite{christakis2008collective,rosenquist2010spread}, or positive, such as generosity  \cite{tsvetkova2014social,pressman2015pif,bisberg2022gift}. In contrast, the socially-induced adoption of communication styles we study in this work is neither positive nor negative and has no ostensible benefits for either the influencer or the follower. It is a harmless communication preference of social networking. Therefore, in the absence of ulterior motives for using Bitmoji stickers, the observations we make here can be explained by either latent homophily -- shared interests that induce both their preference for Bitmoji stickers and their friendship -- or social contagion -- one user's use of Bitmoji stickers socially influenced another user to also use them. Note that these two theories are not mutually exclusive nor conflicting, although quantifying social contagion is challenging when it is usually unavoidably contaminated by network homophily \cite{shalizi2011homophily}. In this work, we present evidence of network homophily, and also present evidence of social contagion while controlling for observable homophilous traits, including their pre-existing tendencies to use Bitmoji stickers and their levels of reciprocity and selectivity in using Bitmoji stickers. It is also possible that explicit social stimuli elicit the behavior of using Bitmoji stickers from users who would want to use them, but were not aware that they existed. We thus argue that both social network effects of homophily and contagion are at play here, and produce the effects we see. Future work could utilize A/B testing to more substantively investigate  how these two theories influence behavior adoption.

\subsubsection{Practical Implications for Platform Designers}
There are several important implications of our work for creators and designers of online social platforms. Some of the most important measures of platform success are user retention and engagement. 
Put broadly, a user's social media presence is influenced by their friends. One practical ramification of this finding is that instead of trying to optimize \textit{individual} user retention, platforms could try to target \textit{collective} user retention based on friend groups. Since social platform engagement is contagious, targeting groups of users could initiate a cascading increase in engagement and retention. 

In addition, we shed light on ways to promote in-app features, either new features or existing ones. As user behavior is contagious, platforms could motivate a few individuals to start using a feature to propagate an organic, ripple adoption of the feature through the social network. For example, relevant Bitmoji stickers suggested via a pop-up based on keywords mentioned in the conversation, prompting a user to use a sticker they otherwise wouldn't have, could spark their friends’ interest in Bitmoji stickers. Such recommendations can be further refined to target users who are known to be more reciprocal or selective, or collectively target groups of users who are more prone to use similar features. This can shave computational costs off by offering recommendations to all users.

\subsubsection{Broader Impact}
This paper is a case study of a specific type of pictorial communication --- Bitmoji stickers --- on Snapchat. That said, our insights can be generalized toward understanding how new formats of pictorial language are adopted and evolved from social interactions. The benefit of limiting our scope to a relatively exclusive communication feature on one social media app is to decouple its social influences from other platforms, something we cannot do with more generic features such as emojis or memes. In this paper, we avoid analyzing unique aspects of Bitmoji stickers but rather their basic functionality as shareable tools of visual communication. Consequently, we believe that the takeaways of this study could help drive research on the adoption of novel, social networking communicative features in the broader CHI community.

\subsection{Limitations and Future Work}
Our work faces several limitations and opens a few new research avenues. First, our study is limited to only one month of data and our choice of time resolutions in RQ3 is fixed to weekly time periods due to patterns of cyclicity. However, with over 100 million sampled users for every analysis, we are confident in the conclusions. Second, our quasi-experimental approach cannot fully account for observable, latent traits of users. As such, we avoid making claims of causality. If permissible, future efforts can be directed toward devising randomized controlled trials on behaviors similar to ours to confirm and validate our findings. Finally, while we show evidence that Bitmoji sticker usage behavior is infectious, we do not know if this process is best described by simple contagion or complex contagion. A simple contagion requires only one exposure for a person to be ``infected'', whereas a complex contagion requires multiple exposures \cite{hodas2014simple,min2018competing}. A closer look into the role of dyadic relationships can also help deepen our understanding of this diffusion process. It might be the case that not all users are equally influential in this process. For example, reciprocal users will adopt the usage of Bitmoji stickers, but only with the friend who initially sent them the sticker, thus terminating the diffusion of behavior. We leave this exploration for future work. 
\subsection{Ethical Statement}
As with any empirical study of human behavior, our analyses  presented risks of distributing and misusing users' private information. To mitigate those risks completely, we took a few protective steps. In accordance with Snap's policies, all user data are deidentified and anonymized by replacing all user IDs with a random ID and then discarding the ID mapping. The data and all analyses are carried out on secure internal cloud storage and servers. No additional details of users, beyond what we described in \S\ref{sec:data}, are collected or stored. Moreover, at no point in this research did we have access to private chat messages, photos, videos, or other types of content. We do not analyze the types or details of the Bitmoji stickers used, only the fact that they \textit{are} used. After the analyses described in this paper were conducted, we also removed the original raw data to comply with Snap's policy.

\section{Conclusion}
In this work, we provide an empirical and quantitative look at how pictorial communication behavior is socially shaped through a case study of how Bitmoji stickers are used on Snapchat. We provide evidence that most Bitmoji sticker users fall on the  spectrum from being reciprocal and selective users to nonreciprocal and unselective (\textbf{RQ1}). That is, the two distinctive characterizations of user behavior -- reciprocity and selectivity -- are correlated. We further find that this behavior obeys theories of network homophily in that friend groups use Bitmoji stickers at similar rates (\textbf{RQ2}). Finally, using a quasi-experimental method, we find evidence of increased Bitmoji sticker usage and also Snapchat engagement in the ensuing weeks after receiving a Bitmoji sticker from a friend (\textbf{RQ3}). Our work carries important implications for a better understanding of social processes involved in communication preferences.
\begin{acks}
\end{acks}
\balance

\bibliographystyle{ACM-Reference-Format}
\bibliography{main}


\begin{thebibliography}{78}


\ifx \showCODEN    \undefined \def \showCODEN     #1{\unskip}     \fi
\ifx \showDOI      \undefined \def \showDOI       #1{#1}\fi
\ifx \showISBNx    \undefined \def \showISBNx     #1{\unskip}     \fi
\ifx \showISBNxiii \undefined \def \showISBNxiii  #1{\unskip}     \fi
\ifx \showISSN     \undefined \def \showISSN      #1{\unskip}     \fi
\ifx \showLCCN     \undefined \def \showLCCN      #1{\unskip}     \fi
\ifx \shownote     \undefined \def \shownote      #1{#1}          \fi
\ifx \showarticletitle \undefined \def \showarticletitle #1{#1}   \fi
\ifx \showURL      \undefined \def \showURL       {\relax}        \fi
\providecommand\bibfield[2]{#2}
\providecommand\bibinfo[2]{#2}
\providecommand\natexlab[1]{#1}
\providecommand\showeprint[2][]{arXiv:#2}

\bibitem[Argyle et~al\mbox{.}(1971)]%
        {argyle1971communication}
\bibfield{author}{\bibinfo{person}{Michael Argyle}, \bibinfo{person}{Florisse
  Alkema}, {and} \bibinfo{person}{Robin Gilmour}.}
  \bibinfo{year}{1971}\natexlab{}.
\newblock \showarticletitle{The communication of friendly and hostile attitudes
  by verbal and non-verbal signals}.
\newblock \bibinfo{journal}{\emph{European Journal of Social Psychology}}
  \bibinfo{volume}{1}, \bibinfo{number}{3} (\bibinfo{year}{1971}),
  \bibinfo{pages}{385--402}.
\newblock


\bibitem[Austin(2011)]%
        {austin2011introduction}
\bibfield{author}{\bibinfo{person}{Peter~C Austin}.}
  \bibinfo{year}{2011}\natexlab{}.
\newblock \showarticletitle{An introduction to propensity score methods for
  reducing the effects of confounding in observational studies}.
\newblock \bibinfo{journal}{\emph{Multivariate Behavioral Research}}
  \bibinfo{volume}{46}, \bibinfo{number}{3} (\bibinfo{year}{2011}),
  \bibinfo{pages}{399--424}.
\newblock


\bibitem[Bajarin(2015)]%
        {vox2015}
\bibfield{author}{\bibinfo{person}{Ben Bajarin}.}
  \bibinfo{year}{2015}\natexlab{}.
\newblock \showarticletitle{The new era of visual communication}.
\newblock \bibinfo{journal}{\emph{Vox}} (\bibinfo{year}{2015}).
\newblock
\urldef\tempurl%
\url{https://www.vox.com/2015/6/16/11563610/the-new-era-of-visual-communication}
\showURL{%
Retrieved Aug 22, 2022 from \tempurl}


\bibitem[Barsade(2002)]%
        {barsade2002ripple}
\bibfield{author}{\bibinfo{person}{Sigal~G Barsade}.}
  \bibinfo{year}{2002}\natexlab{}.
\newblock \showarticletitle{The ripple effect: Emotional contagion and its
  influence on group behavior}.
\newblock \bibinfo{journal}{\emph{Administrative Science Quarterly}}
  \bibinfo{volume}{47}, \bibinfo{number}{4} (\bibinfo{year}{2002}),
  \bibinfo{pages}{644--675}.
\newblock


\bibitem[Bastiampillai et~al\mbox{.}(2013)]%
        {bastiampillai2013depression}
\bibfield{author}{\bibinfo{person}{Tarun Bastiampillai},
  \bibinfo{person}{Stephen Allison}, {and} \bibinfo{person}{Sherry Chan}.}
  \bibinfo{year}{2013}\natexlab{}.
\newblock \showarticletitle{Is depression contagious? The importance of social
  networks and the implications of contagion theory}.
\newblock \bibinfo{journal}{\emph{Australian \& New Zealand Journal of
  Psychiatry}} \bibinfo{volume}{47}, \bibinfo{number}{4}
  (\bibinfo{year}{2013}), \bibinfo{pages}{299--303}.
\newblock


\bibitem[Billy et~al\mbox{.}(1984)]%
        {billy1984adolescent}
\bibfield{author}{\bibinfo{person}{John~OG Billy}, \bibinfo{person}{J~Richard
  Udry}, {and} \bibinfo{person}{Joseph~Lee Rodgers}.}
  \bibinfo{year}{1984}\natexlab{}.
\newblock \showarticletitle{Adolescent sexual behavior and friendship choice}.
\newblock \bibinfo{journal}{\emph{Social Forces}} \bibinfo{volume}{62},
  \bibinfo{number}{3} (\bibinfo{year}{1984}), \bibinfo{pages}{653--678}.
\newblock


\bibitem[Bisberg et~al\mbox{.}(2022)]%
        {bisberg2022gift}
\bibfield{author}{\bibinfo{person}{Alexander~J Bisberg}, \bibinfo{person}{Julie
  Jiang}, \bibinfo{person}{Yilei Zeng}, \bibinfo{person}{Emily Chen}, {and}
  \bibinfo{person}{Emilio Ferrara}.} \bibinfo{year}{2022}\natexlab{}.
\newblock \showarticletitle{The gift that keeps on giving: Generosity is
  contagious in multiplayer online games}, In \bibinfo{booktitle}{CSCW '22}.
\newblock \bibinfo{journal}{\emph{PACM HCI}} \bibinfo{volume}{6},
  \bibinfo{number}{CSCW2}, Article \bibinfo{articleno}{395},
  \bibinfo{numpages}{22}~pages.
\newblock
\urldef\tempurl%
\url{https://doi.org/10.1145/3555120}
\showDOI{\tempurl}


\bibitem[Bisgin et~al\mbox{.}(2012)]%
        {bisgin2012study}
\bibfield{author}{\bibinfo{person}{Halil Bisgin}, \bibinfo{person}{Nitin
  Agarwal}, {and} \bibinfo{person}{Xiaowei Xu}.}
  \bibinfo{year}{2012}\natexlab{}.
\newblock \showarticletitle{A study of homophily on social media}.
\newblock \bibinfo{journal}{\emph{World Wide Web}} \bibinfo{volume}{15},
  \bibinfo{number}{2} (\bibinfo{year}{2012}), \bibinfo{pages}{213--232}.
\newblock


\bibitem[{Bitmoji Support}(2021)]%
        {bitmojicustomize}
\bibfield{author}{\bibinfo{person}{{Bitmoji Support}}.}
  \bibinfo{year}{2021}\natexlab{}.
\newblock \showarticletitle{Customize text on Bitmoji stickers}.
\newblock \bibinfo{journal}{\emph{Bitmoji Support}} (\bibinfo{year}{2021}).
\newblock
\urldef\tempurl%
\url{https://support.bitmoji.com/hc/en-us/articles/360034632291-Customize-Text-on-Bitmoji-Stickers}
\showURL{%
Retrieved Aug 22, 2022 from \tempurl}


\bibitem[{Bitmoji Support}(2022)]%
        {bitmojideluxe}
\bibfield{author}{\bibinfo{person}{{Bitmoji Support}}.}
  \bibinfo{year}{2022}\natexlab{}.
\newblock \showarticletitle{Bitmoji deluxe}.
\newblock \bibinfo{journal}{\emph{Bitmoji Support}} (\bibinfo{year}{2022}).
\newblock
\urldef\tempurl%
\url{https://support.bitmoji.com/hc/en-us/articles/360025043212-Bitmoji-Deluxe}
\showURL{%
Retrieved Aug 22, 2022 from \tempurl}


\bibitem[Byrne and Griffith(1973)]%
        {byrne1973interpersonal}
\bibfield{author}{\bibinfo{person}{Donn Byrne} {and} \bibinfo{person}{William
  Griffith}.} \bibinfo{year}{1973}\natexlab{}.
\newblock \showarticletitle{Interpersonal attraction}.
\newblock \bibinfo{journal}{\emph{Annual Review of Psychology}}
  (\bibinfo{year}{1973}).
\newblock


\bibitem[Byron and Baldridge(2007)]%
        {byron2007mail}
\bibfield{author}{\bibinfo{person}{Kristin Byron} {and}
  \bibinfo{person}{David~C Baldridge}.} \bibinfo{year}{2007}\natexlab{}.
\newblock \showarticletitle{E-mail recipients' impressions of senders'
  likability: The interactive effect of nonverbal cues and recipients'
  personality}.
\newblock \bibinfo{journal}{\emph{The Journal of Business Communication
  (1973)}} \bibinfo{volume}{44}, \bibinfo{number}{2} (\bibinfo{year}{2007}),
  \bibinfo{pages}{137--160}.
\newblock


\bibitem[Carstensen et~al\mbox{.}(1999)]%
        {carstensen1999taking}
\bibfield{author}{\bibinfo{person}{Laura~L Carstensen},
  \bibinfo{person}{Derek~M Isaacowitz}, {and} \bibinfo{person}{Susan~T
  Charles}.} \bibinfo{year}{1999}\natexlab{}.
\newblock \showarticletitle{Taking time seriously: A theory of socioemotional
  selectivity.}
\newblock \bibinfo{journal}{\emph{American Psychologist}} \bibinfo{volume}{54},
  \bibinfo{number}{3} (\bibinfo{year}{1999}), \bibinfo{pages}{165}.
\newblock


\bibitem[Centola(2011)]%
        {centola2011experimental}
\bibfield{author}{\bibinfo{person}{Damon Centola}.}
  \bibinfo{year}{2011}\natexlab{}.
\newblock \showarticletitle{An experimental study of homophily in the adoption
  of health behavior}.
\newblock \bibinfo{journal}{\emph{Science}} \bibinfo{volume}{334},
  \bibinfo{number}{6060} (\bibinfo{year}{2011}), \bibinfo{pages}{1269--1272}.
\newblock


\bibitem[Christakis and Fowler(2008)]%
        {christakis2008collective}
\bibfield{author}{\bibinfo{person}{Nicholas~A Christakis} {and}
  \bibinfo{person}{James~H Fowler}.} \bibinfo{year}{2008}\natexlab{}.
\newblock \showarticletitle{The collective dynamics of smoking in a large
  social network}.
\newblock \bibinfo{journal}{\emph{New England Journal of Medicine}}
  \bibinfo{volume}{358}, \bibinfo{number}{21} (\bibinfo{year}{2008}),
  \bibinfo{pages}{2249--2258}.
\newblock


\bibitem[Christakis and Fowler(2013)]%
        {christakis2013social}
\bibfield{author}{\bibinfo{person}{Nicholas~A Christakis} {and}
  \bibinfo{person}{James~H Fowler}.} \bibinfo{year}{2013}\natexlab{}.
\newblock \showarticletitle{Social contagion theory: Examining dynamic social
  networks and human behavior}.
\newblock \bibinfo{journal}{\emph{Statistics in Medicine}}
  \bibinfo{volume}{32}, \bibinfo{number}{4} (\bibinfo{year}{2013}),
  \bibinfo{pages}{556--577}.
\newblock


\bibitem[Cohen(1977)]%
        {cohen1977sources}
\bibfield{author}{\bibinfo{person}{Jere~M Cohen}.}
  \bibinfo{year}{1977}\natexlab{}.
\newblock \showarticletitle{Sources of peer group homogeneity}.
\newblock \bibinfo{journal}{\emph{Sociology of Education}}
  (\bibinfo{year}{1977}), \bibinfo{pages}{227--241}.
\newblock


\bibitem[De~Choudhury et~al\mbox{.}(2016)]%
        {de2016discovering}
\bibfield{author}{\bibinfo{person}{Munmun De~Choudhury}, \bibinfo{person}{Emre
  Kiciman}, \bibinfo{person}{Mark Dredze}, \bibinfo{person}{Glen Coppersmith},
  {and} \bibinfo{person}{Mrinal Kumar}.} \bibinfo{year}{2016}\natexlab{}.
\newblock \showarticletitle{Discovering shifts to suicidal ideation from mental
  health content in social media}. In \bibinfo{booktitle}{\emph{CHI '16}}.
  \bibinfo{pages}{2098--2110}.
\newblock


\bibitem[De~Seta(2018)]%
        {de2018biaoqing}
\bibfield{author}{\bibinfo{person}{Gabriele De~Seta}.}
  \bibinfo{year}{2018}\natexlab{}.
\newblock \showarticletitle{Biaoqing: The circulation of emoticons, emoji,
  stickers, and custom images on Chinese digital media platforms}.
\newblock \bibinfo{journal}{\emph{First Monday}} (\bibinfo{year}{2018}).
\newblock


\bibitem[Dubey et~al\mbox{.}(2021)]%
        {dubey2021curiosity}
\bibfield{author}{\bibinfo{person}{Rachit Dubey}, \bibinfo{person}{Hermish
  Mehta}, {and} \bibinfo{person}{Tania Lombrozo}.}
  \bibinfo{year}{2021}\natexlab{}.
\newblock \showarticletitle{Curiosity is contagious: A social influence
  intervention to induce curiosity}.
\newblock \bibinfo{journal}{\emph{Cognitive Science}} \bibinfo{volume}{45},
  \bibinfo{number}{2} (\bibinfo{year}{2021}), \bibinfo{pages}{e12937}.
\newblock


\bibitem[Duck(1973)]%
        {duck1973personal}
\bibfield{author}{\bibinfo{person}{Steven~W Duck}.}
  \bibinfo{year}{1973}\natexlab{}.
\newblock \bibinfo{booktitle}{\emph{Personal relationships and personal
  constructs: A study of friendship formation}}.
\newblock \bibinfo{publisher}{John Wiley \& Sons}.
\newblock


\bibitem[Eastwick and Finkel(2009)]%
        {eastwick2009reciprocity}
\bibfield{author}{\bibinfo{person}{Paul~W Eastwick} {and}
  \bibinfo{person}{Eli~J Finkel}.} \bibinfo{year}{2009}\natexlab{}.
\newblock \showarticletitle{Reciprocity of liking}.
\newblock In \bibinfo{booktitle}{\emph{Encyclopedia of Human Relationships}}.
  \bibinfo{publisher}{SAGE Publications, Inc}, \bibinfo{pages}{1333--1336}.
\newblock


\bibitem[Eastwick et~al\mbox{.}(2007)]%
        {eastwick2007selective}
\bibfield{author}{\bibinfo{person}{Paul~W Eastwick}, \bibinfo{person}{Eli~J
  Finkel}, \bibinfo{person}{Daniel Mochon}, {and} \bibinfo{person}{Dan
  Ariely}.} \bibinfo{year}{2007}\natexlab{}.
\newblock \showarticletitle{Selective versus unselective romantic desire: Not
  all reciprocity is created equal}.
\newblock \bibinfo{journal}{\emph{Psychological Science}} \bibinfo{volume}{18},
  \bibinfo{number}{4} (\bibinfo{year}{2007}), \bibinfo{pages}{317--319}.
\newblock


\bibitem[Fehr and G{\"a}chter(2000)]%
        {fehr2000fairness}
\bibfield{author}{\bibinfo{person}{Ernst Fehr} {and} \bibinfo{person}{Simon
  G{\"a}chter}.} \bibinfo{year}{2000}\natexlab{}.
\newblock \showarticletitle{Fairness and retaliation: The economics of
  reciprocity}.
\newblock \bibinfo{journal}{\emph{Journal of Economic Perspectives}}
  \bibinfo{volume}{14}, \bibinfo{number}{3} (\bibinfo{year}{2000}),
  \bibinfo{pages}{159--181}.
\newblock


\bibitem[Ferrara and Yang(2015a)]%
        {ferrara2015measuring}
\bibfield{author}{\bibinfo{person}{Emilio Ferrara} {and} \bibinfo{person}{Zeyao
  Yang}.} \bibinfo{year}{2015}\natexlab{a}.
\newblock \showarticletitle{Measuring emotional contagion in social media}.
\newblock \bibinfo{journal}{\emph{PloS One}} \bibinfo{volume}{10},
  \bibinfo{number}{11} (\bibinfo{year}{2015}), \bibinfo{pages}{e0142390}.
\newblock


\bibitem[Ferrara and Yang(2015b)]%
        {ferrara2015quantifying}
\bibfield{author}{\bibinfo{person}{Emilio Ferrara} {and} \bibinfo{person}{Zeyao
  Yang}.} \bibinfo{year}{2015}\natexlab{b}.
\newblock \showarticletitle{Quantifying the effect of sentiment on information
  diffusion in social media}.
\newblock \bibinfo{journal}{\emph{PeerJ Computer Science}}  \bibinfo{volume}{1}
  (\bibinfo{year}{2015}), \bibinfo{pages}{e26}.
\newblock


\bibitem[Figeac and Favre(2021)]%
        {figeac2021behavioral}
\bibfield{author}{\bibinfo{person}{Julien Figeac} {and}
  \bibinfo{person}{Guillaume Favre}.} \bibinfo{year}{2021}\natexlab{}.
\newblock \showarticletitle{How behavioral homophily on social media influences
  the perception of tie-strengthening within young adults’ personal
  networks}.
\newblock \bibinfo{journal}{\emph{New Media \& Society}}
  (\bibinfo{year}{2021}), \bibinfo{pages}{14614448211020691}.
\newblock


\bibitem[Forristal(2022)]%
        {techcrunch2022}
\bibfield{author}{\bibinfo{person}{Lauren Forristal}.}
  \bibinfo{year}{2022}\natexlab{}.
\newblock \showarticletitle{Game studio HiDef partners with Snap to develop a
  Bitmoji dance social mobile game}.
\newblock \bibinfo{journal}{\emph{TechCrunch}} (\bibinfo{year}{2022}).
\newblock
\urldef\tempurl%
\url{https://techcrunch.com/2022/06/07/game-studio-hidef-partners-with-snap-to-develop-a-bitmoji-dance-social-mobile-game/}
\showURL{%
Retrieved Sep 10, 2022 from \tempurl}


\bibitem[Gelman and Hill(2006)]%
        {gelman2007causal}
\bibfield{author}{\bibinfo{person}{Andrew Gelman} {and}
  \bibinfo{person}{Jennifer Hill}.} \bibinfo{year}{2006}\natexlab{}.
\newblock \bibinfo{booktitle}{\emph{Data analysis using regression and
  multilevel/hierarchical models}}.
\newblock \bibinfo{publisher}{Cambridge University Oress}.
\newblock


\bibitem[Gesselman et~al\mbox{.}(2019)]%
        {gesselman2019worth}
\bibfield{author}{\bibinfo{person}{Amanda~N Gesselman},
  \bibinfo{person}{Vivian~P Ta}, {and} \bibinfo{person}{Justin~R Garcia}.}
  \bibinfo{year}{2019}\natexlab{}.
\newblock \showarticletitle{Worth a thousand interpersonal words: Emoji as
  affective signals for relationship-oriented digital communication}.
\newblock \bibinfo{journal}{\emph{PloS One}} \bibinfo{volume}{14},
  \bibinfo{number}{8} (\bibinfo{year}{2019}), \bibinfo{pages}{e0221297}.
\newblock


\bibitem[Gintis(2000)]%
        {gintis2000strong}
\bibfield{author}{\bibinfo{person}{Herbert Gintis}.}
  \bibinfo{year}{2000}\natexlab{}.
\newblock \showarticletitle{Strong reciprocity and human sociality}.
\newblock \bibinfo{journal}{\emph{Journal of Theoretical Biology}}
  \bibinfo{volume}{206}, \bibinfo{number}{2} (\bibinfo{year}{2000}),
  \bibinfo{pages}{169--179}.
\newblock


\bibitem[Hallinan(1978)]%
        {hallinan1978process}
\bibfield{author}{\bibinfo{person}{Maureen~T Hallinan}.}
  \bibinfo{year}{1978}\natexlab{}.
\newblock \showarticletitle{The process of friendship formation}.
\newblock \bibinfo{journal}{\emph{Social Networks}} \bibinfo{volume}{1},
  \bibinfo{number}{2} (\bibinfo{year}{1978}), \bibinfo{pages}{193--210}.
\newblock


\bibitem[Hariton and Locascio(2018)]%
        {hariton2018randomised}
\bibfield{author}{\bibinfo{person}{Eduardo Hariton} {and}
  \bibinfo{person}{Joseph~J Locascio}.} \bibinfo{year}{2018}\natexlab{}.
\newblock \showarticletitle{Randomised controlled trials—the gold standard
  for effectiveness research}.
\newblock \bibinfo{journal}{\emph{BJOG: An International Journal of Obstetrics
  and Gynaecology}} \bibinfo{volume}{125}, \bibinfo{number}{13}
  (\bibinfo{year}{2018}), \bibinfo{pages}{1716}.
\newblock


\bibitem[Hodas and Lerman(2014)]%
        {hodas2014simple}
\bibfield{author}{\bibinfo{person}{Nathan~O Hodas} {and}
  \bibinfo{person}{Kristina Lerman}.} \bibinfo{year}{2014}\natexlab{}.
\newblock \showarticletitle{The simple rules of social contagion}.
\newblock \bibinfo{journal}{\emph{Scientific Reports}} \bibinfo{volume}{4},
  \bibinfo{number}{1} (\bibinfo{year}{2014}), \bibinfo{pages}{1--7}.
\newblock


\bibitem[Huston and Levinger(1978)]%
        {huston1978interpersonal}
\bibfield{author}{\bibinfo{person}{Ted~L Huston} {and} \bibinfo{person}{George
  Levinger}.} \bibinfo{year}{1978}\natexlab{}.
\newblock \showarticletitle{Interpersonal attraction and relationships}.
\newblock \bibinfo{journal}{\emph{Annual Review of Psychology}}
  \bibinfo{volume}{29}, \bibinfo{number}{1} (\bibinfo{year}{1978}),
  \bibinfo{pages}{115--156}.
\newblock


\bibitem[Imbens and Rubin(2015)]%
        {imbens2015causal}
\bibfield{author}{\bibinfo{person}{Guido~W Imbens} {and}
  \bibinfo{person}{Donald~B Rubin}.} \bibinfo{year}{2015}\natexlab{}.
\newblock \bibinfo{booktitle}{\emph{Causal inference in statistics, social, and
  biomedical sciences}}.
\newblock \bibinfo{publisher}{Cambridge University Press}.
\newblock


\bibitem[Janssen et~al\mbox{.}(2014)]%
        {janssen2014affective}
\bibfield{author}{\bibinfo{person}{Joris~H Janssen}, \bibinfo{person}{Wijnand~A
  Ijsselsteijn}, {and} \bibinfo{person}{Joyce~HDM Westerink}.}
  \bibinfo{year}{2014}\natexlab{}.
\newblock \showarticletitle{How affective technologies can influence intimate
  interactions and improve social connectedness}.
\newblock \bibinfo{journal}{\emph{International Journal of Human-Computer
  Studies}} \bibinfo{volume}{72}, \bibinfo{number}{1} (\bibinfo{year}{2014}),
  \bibinfo{pages}{33--43}.
\newblock


\bibitem[Jiang et~al\mbox{.}(2018)]%
        {jiang2018perfect}
\bibfield{author}{\bibinfo{person}{Jialun~``Aaron'' Jiang},
  \bibinfo{person}{Casey Fiesler}, {and} \bibinfo{person}{Jed~R Brubaker}.}
  \bibinfo{year}{2018}\natexlab{}.
\newblock \showarticletitle{``The Perfect One'': Understanding communication
  practices and challenges with animated GIFs}.
\newblock \bibinfo{journal}{\emph{PACM HCI}} \bibinfo{volume}{2},
  \bibinfo{number}{CSCW} (\bibinfo{year}{2018}), \bibinfo{pages}{1--20}.
\newblock


\bibitem[Kandel(1978)]%
        {kandel1978homophily}
\bibfield{author}{\bibinfo{person}{Denise~B Kandel}.}
  \bibinfo{year}{1978}\natexlab{}.
\newblock \showarticletitle{Homophily, Selection, and Socialization in
  Adolescent Friendships 1}.
\newblock \bibinfo{journal}{\emph{American Journal of Sociology}}
  \bibinfo{volume}{84}, \bibinfo{number}{2} (\bibinfo{year}{1978}),
  \bibinfo{pages}{427--436}.
\newblock


\bibitem[King and Nielsen(2019)]%
        {king2019propensity}
\bibfield{author}{\bibinfo{person}{Gary King} {and}
  \bibinfo{person}{Richard~Alexander Nielsen}.}
  \bibinfo{year}{2019}\natexlab{}.
\newblock \showarticletitle{Why propensity scores should not be used for
  matching}.
\newblock \bibinfo{journal}{\emph{Political Analysis}}  \bibinfo{volume}{27}
  (\bibinfo{year}{2019}), \bibinfo{pages}{435--454}.
\newblock
\urldef\tempurl%
\url{https://doi.org/10.1017/pan.2019.11}
\showDOI{\tempurl}


\bibitem[Kingma and Ba(2015)]%
        {kingma2014adam}
\bibfield{author}{\bibinfo{person}{Diederik~P Kingma} {and}
  \bibinfo{person}{Jimmy Ba}.} \bibinfo{year}{2015}\natexlab{}.
\newblock \showarticletitle{Adam: A method for stochastic optimization}.
\newblock \bibinfo{journal}{\emph{ICLR '15}}.
\newblock


\bibitem[Knoke(1990)]%
        {knoke1990networks}
\bibfield{author}{\bibinfo{person}{David Knoke}.}
  \bibinfo{year}{1990}\natexlab{}.
\newblock \showarticletitle{Networks of political action: Toward theory
  construction}.
\newblock \bibinfo{journal}{\emph{Social Forces}} \bibinfo{volume}{68},
  \bibinfo{number}{4} (\bibinfo{year}{1990}), \bibinfo{pages}{1041--1063}.
\newblock


\bibitem[Kohavi and Longbotham(2017)]%
        {kohavi2017online}
\bibfield{author}{\bibinfo{person}{Ron Kohavi} {and} \bibinfo{person}{Roger
  Longbotham}.} \bibinfo{year}{2017}\natexlab{}.
\newblock \showarticletitle{Online controlled experiments and A/B testing.}
\newblock \bibinfo{journal}{\emph{Encyclopedia of Machine Learning and Data
  Mining}} \bibinfo{volume}{7}, \bibinfo{number}{8} (\bibinfo{year}{2017}),
  \bibinfo{pages}{922--929}.
\newblock


\bibitem[Kossinets and Watts(2009)]%
        {kossinets2009origins}
\bibfield{author}{\bibinfo{person}{Gueorgi Kossinets} {and}
  \bibinfo{person}{Duncan~J Watts}.} \bibinfo{year}{2009}\natexlab{}.
\newblock \showarticletitle{Origins of homophily in an evolving social
  network}.
\newblock \bibinfo{journal}{\emph{American Journal of Sociology}}
  \bibinfo{volume}{115}, \bibinfo{number}{2} (\bibinfo{year}{2009}),
  \bibinfo{pages}{405--450}.
\newblock


\bibitem[Kramer et~al\mbox{.}(2014)]%
        {kramer2014experimental}
\bibfield{author}{\bibinfo{person}{Adam~DI Kramer}, \bibinfo{person}{Jamie~E
  Guillory}, {and} \bibinfo{person}{Jeffrey~T Hancock}.}
  \bibinfo{year}{2014}\natexlab{}.
\newblock \showarticletitle{Experimental evidence of massive-scale emotional
  contagion through social networks}.
\newblock \bibinfo{journal}{\emph{Proceedings of the National Academy of
  Sciences}} \bibinfo{volume}{111}, \bibinfo{number}{24}
  (\bibinfo{year}{2014}), \bibinfo{pages}{8788--8790}.
\newblock


\bibitem[Lacoma and Beaton(2021)]%
        {digital2021bitmoji}
\bibfield{author}{\bibinfo{person}{Tyler Lacoma} {and} \bibinfo{person}{Paula
  Beaton}.} \bibinfo{year}{2021}\natexlab{}.
\newblock \showarticletitle{What is Bitmoji? Everything you need to know}.
\newblock \bibinfo{journal}{\emph{Digital Trends}} (\bibinfo{year}{2021}).
\newblock
\urldef\tempurl%
\url{https://www.digitaltrends.com/mobile/what-is-bitmoji/}
\showURL{%
Retrieved Aug 22, 2022 from \tempurl}


\bibitem[Lo(2008)]%
        {lo2008nonverbal}
\bibfield{author}{\bibinfo{person}{Shao-Kang Lo}.}
  \bibinfo{year}{2008}\natexlab{}.
\newblock \showarticletitle{The nonverbal communication functions of emoticons
  in computer-mediated communication}.
\newblock \bibinfo{journal}{\emph{Cyberpsychology \& Behavior}}
  \bibinfo{volume}{11}, \bibinfo{number}{5} (\bibinfo{year}{2008}),
  \bibinfo{pages}{595--597}.
\newblock


\bibitem[Maldeniya et~al\mbox{.}(2020)]%
        {maldeniya2020herding}
\bibfield{author}{\bibinfo{person}{Danaja Maldeniya}, \bibinfo{person}{Ceren
  Budak}, \bibinfo{person}{Lionel~P Robert~Jr}, {and} \bibinfo{person}{Daniel~M
  Romero}.} \bibinfo{year}{2020}\natexlab{}.
\newblock \showarticletitle{Herding a deluge of good samaritans: How GitHub
  projects respond to increased attention}. In
  \bibinfo{booktitle}{\emph{Proceedings of The Web Conference 2020}}.
  \bibinfo{pages}{2055--2065}.
\newblock


\bibitem[Marsden(1998)]%
        {marsden1998memetics}
\bibfield{author}{\bibinfo{person}{Paul Marsden}.}
  \bibinfo{year}{1998}\natexlab{}.
\newblock \showarticletitle{Memetics and social contagion: Two sides of the
  same coin}.
\newblock \bibinfo{journal}{\emph{Journal of Memetics-Evolutionary Models of
  Information Transmission}} \bibinfo{volume}{2}, \bibinfo{number}{2}
  (\bibinfo{year}{1998}), \bibinfo{pages}{171--185}.
\newblock


\bibitem[McCarney et~al\mbox{.}(2007)]%
        {mccarney2007hawthorne}
\bibfield{author}{\bibinfo{person}{Rob McCarney}, \bibinfo{person}{James
  Warner}, \bibinfo{person}{Steve Iliffe}, \bibinfo{person}{Robbert
  Van~Haselen}, \bibinfo{person}{Mark Griffin}, {and} \bibinfo{person}{Peter
  Fisher}.} \bibinfo{year}{2007}\natexlab{}.
\newblock \showarticletitle{The Hawthorne Effect: a randomised, controlled
  trial}.
\newblock \bibinfo{journal}{\emph{BMC Medical Research Methodology}}
  \bibinfo{volume}{7}, \bibinfo{number}{1} (\bibinfo{year}{2007}),
  \bibinfo{pages}{1--8}.
\newblock


\bibitem[McPherson et~al\mbox{.}(2001)]%
        {mcpherson2001birds}
\bibfield{author}{\bibinfo{person}{Miller McPherson}, \bibinfo{person}{Lynn
  Smith-Lovin}, {and} \bibinfo{person}{James~M Cook}.}
  \bibinfo{year}{2001}\natexlab{}.
\newblock \showarticletitle{Birds of a feather: Homophily in social networks}.
\newblock \bibinfo{journal}{\emph{Annual Review of Sociology}}
  (\bibinfo{year}{2001}), \bibinfo{pages}{415--444}.
\newblock


\bibitem[Min and San~Miguel(2018)]%
        {min2018competing}
\bibfield{author}{\bibinfo{person}{Byungjoon Min} {and} \bibinfo{person}{Maxi
  San~Miguel}.} \bibinfo{year}{2018}\natexlab{}.
\newblock \showarticletitle{Competing contagion processes: Complex contagion
  triggered by simple contagion}.
\newblock \bibinfo{journal}{\emph{Scientific Reports}} \bibinfo{volume}{8},
  \bibinfo{number}{1} (\bibinfo{year}{2018}), \bibinfo{pages}{1--8}.
\newblock


\bibitem[M{\o}nsted et~al\mbox{.}(2017)]%
        {monsted2017evidence}
\bibfield{author}{\bibinfo{person}{Bjarke M{\o}nsted}, \bibinfo{person}{Piotr
  Sapie{\.z}y{\'n}ski}, \bibinfo{person}{Emilio Ferrara}, {and}
  \bibinfo{person}{Sune Lehmann}.} \bibinfo{year}{2017}\natexlab{}.
\newblock \showarticletitle{Evidence of complex contagion of information in
  social media: An experiment using Twitter bots}.
\newblock \bibinfo{journal}{\emph{PloS One}} \bibinfo{volume}{12},
  \bibinfo{number}{9} (\bibinfo{year}{2017}), \bibinfo{pages}{e0184148}.
\newblock


\bibitem[Newman(2002)]%
        {newman2002assortative}
\bibfield{author}{\bibinfo{person}{Mark~EJ Newman}.}
  \bibinfo{year}{2002}\natexlab{}.
\newblock \showarticletitle{Assortative mixing in networks}.
\newblock \bibinfo{journal}{\emph{Physical Review Letters}}
  \bibinfo{volume}{89}, \bibinfo{number}{20} (\bibinfo{year}{2002}),
  \bibinfo{pages}{208701}.
\newblock


\bibitem[Newman(2003)]%
        {newman2003mixing}
\bibfield{author}{\bibinfo{person}{Mark~EJ Newman}.}
  \bibinfo{year}{2003}\natexlab{}.
\newblock \showarticletitle{Mixing patterns in networks}.
\newblock \bibinfo{journal}{\emph{Physical Review E}} \bibinfo{volume}{67},
  \bibinfo{number}{2} (\bibinfo{year}{2003}), \bibinfo{pages}{026126}.
\newblock


\bibitem[Olteanu et~al\mbox{.}(2017)]%
        {olteanu2017distilling}
\bibfield{author}{\bibinfo{person}{Alexandra Olteanu}, \bibinfo{person}{Onur
  Varol}, {and} \bibinfo{person}{Emre Kiciman}.}
  \bibinfo{year}{2017}\natexlab{}.
\newblock \showarticletitle{Distilling the outcomes of personal experiences: A
  propensity-scored analysis of social media}. In
  \bibinfo{booktitle}{\emph{CSCW '17}}. \bibinfo{pages}{370--386}.
\newblock


\bibitem[Pachucki et~al\mbox{.}(2011)]%
        {pachucki2011social}
\bibfield{author}{\bibinfo{person}{Mark~A Pachucki}, \bibinfo{person}{Paul~F
  Jacques}, {and} \bibinfo{person}{Nicholas~A Christakis}.}
  \bibinfo{year}{2011}\natexlab{}.
\newblock \showarticletitle{Social network concordance in food choice among
  spouses, friends, and siblings}.
\newblock \bibinfo{journal}{\emph{American Journal of Public Health}}
  \bibinfo{volume}{101}, \bibinfo{number}{11} (\bibinfo{year}{2011}),
  \bibinfo{pages}{2170--2177}.
\newblock


\bibitem[Pressman et~al\mbox{.}(2015)]%
        {pressman2015pif}
\bibfield{author}{\bibinfo{person}{Sarah~D Pressman}, \bibinfo{person}{Tara~L
  Kraft}, {and} \bibinfo{person}{Marie~P Cross}.}
  \bibinfo{year}{2015}\natexlab{}.
\newblock \showarticletitle{It’s good to do good and receive good: The impact
  of a ‘pay it forward’style kindness intervention on giver and receiver
  well-being}.
\newblock \bibinfo{journal}{\emph{The Journal of Positive Psychology}}
  \bibinfo{volume}{10}, \bibinfo{number}{4} (\bibinfo{year}{2015}),
  \bibinfo{pages}{293--302}.
\newblock
\urldef\tempurl%
\url{https://doi.org/10.1080/17439760.2014.965269}
\showDOI{\tempurl}


\bibitem[Rosenbaum and Rubin(1984)]%
        {rosenbaum1984reducing}
\bibfield{author}{\bibinfo{person}{Paul~R Rosenbaum} {and}
  \bibinfo{person}{Donald~B Rubin}.} \bibinfo{year}{1984}\natexlab{}.
\newblock \showarticletitle{Reducing bias in observational studies using
  subclassification on the propensity score}.
\newblock \bibinfo{journal}{\emph{J. Amer. Statist. Assoc.}}
  \bibinfo{volume}{79}, \bibinfo{number}{387} (\bibinfo{year}{1984}),
  \bibinfo{pages}{516--524}.
\newblock


\bibitem[Rosenquist et~al\mbox{.}(2011)]%
        {rosenquist2011social}
\bibfield{author}{\bibinfo{person}{J~Niels Rosenquist},
  \bibinfo{person}{James~H Fowler}, {and} \bibinfo{person}{Nicholas~A
  Christakis}.} \bibinfo{year}{2011}\natexlab{}.
\newblock \showarticletitle{Social network determinants of depression}.
\newblock \bibinfo{journal}{\emph{Molecular Psychiatry}} \bibinfo{volume}{16},
  \bibinfo{number}{3} (\bibinfo{year}{2011}), \bibinfo{pages}{273--281}.
\newblock


\bibitem[Rosenquist et~al\mbox{.}(2010)]%
        {rosenquist2010spread}
\bibfield{author}{\bibinfo{person}{J~Niels Rosenquist}, \bibinfo{person}{Joanne
  Murabito}, \bibinfo{person}{James~H Fowler}, {and}
  \bibinfo{person}{Nicholas~A Christakis}.} \bibinfo{year}{2010}\natexlab{}.
\newblock \showarticletitle{The spread of alcohol consumption behavior in a
  large social network}.
\newblock \bibinfo{journal}{\emph{Annals of Internal Medicine}}
  \bibinfo{volume}{152}, \bibinfo{number}{7} (\bibinfo{year}{2010}),
  \bibinfo{pages}{426--433}.
\newblock


\bibitem[Saha et~al\mbox{.}(2021)]%
        {saha2021advertiming}
\bibfield{author}{\bibinfo{person}{Koustuv Saha}, \bibinfo{person}{Yozen Liu},
  \bibinfo{person}{Nicholas Vincent}, \bibinfo{person}{Farhan~Asif Chowdhury},
  \bibinfo{person}{Leonardo Neves}, \bibinfo{person}{Neil Shah}, {and}
  \bibinfo{person}{Maarten~W Bos}.} \bibinfo{year}{2021}\natexlab{}.
\newblock \showarticletitle{Advertiming matters: Examining user ad consumption
  for effective ad allocations on social media}. In
  \bibinfo{booktitle}{\emph{CHI '21}} (Yokohama, Japan).
  \bibinfo{publisher}{ACM}, Article \bibinfo{articleno}{581},
  \bibinfo{numpages}{18}~pages.
\newblock
\urldef\tempurl%
\url{https://doi.org/10.1145/3411764.3445394}
\showDOI{\tempurl}


\bibitem[Saha et~al\mbox{.}(2019)]%
        {saha2019social}
\bibfield{author}{\bibinfo{person}{Koustuv Saha}, \bibinfo{person}{Benjamin
  Sugar}, \bibinfo{person}{John Torous}, \bibinfo{person}{Bruno Abrahao},
  \bibinfo{person}{Emre K{\i}c{\i}man}, {and} \bibinfo{person}{Munmun
  De~Choudhury}.} \bibinfo{year}{2019}\natexlab{}.
\newblock \showarticletitle{A social media study on the effects of psychiatric
  medication use}. In \bibinfo{booktitle}{\emph{ICWSM '13}},
  Vol.~\bibinfo{volume}{13}. \bibinfo{pages}{440--451}.
\newblock


\bibitem[Setoguchi et~al\mbox{.}(2008)]%
        {setoguchi2008evaluating}
\bibfield{author}{\bibinfo{person}{Soko Setoguchi}, \bibinfo{person}{Sebastian
  Schneeweiss}, \bibinfo{person}{M~Alan Brookhart}, \bibinfo{person}{Robert~J
  Glynn}, {and} \bibinfo{person}{E~Francis Cook}.}
  \bibinfo{year}{2008}\natexlab{}.
\newblock \showarticletitle{Evaluating uses of data mining techniques in
  propensity score estimation: a simulation study}.
\newblock \bibinfo{journal}{\emph{Pharmacoepidemiology and Drug Safety}}
  \bibinfo{volume}{17}, \bibinfo{number}{6} (\bibinfo{year}{2008}),
  \bibinfo{pages}{546--555}.
\newblock


\bibitem[Shalizi and Thomas(2011)]%
        {shalizi2011homophily}
\bibfield{author}{\bibinfo{person}{Cosma~Rohilla Shalizi} {and}
  \bibinfo{person}{Andrew~C Thomas}.} \bibinfo{year}{2011}\natexlab{}.
\newblock \showarticletitle{Homophily and contagion are generically confounded
  in observational social network studies}.
\newblock \bibinfo{journal}{\emph{Sociological methods \& research}}
  \bibinfo{volume}{40}, \bibinfo{number}{2} (\bibinfo{year}{2011}),
  \bibinfo{pages}{211--239}.
\newblock


\bibitem[Shandilya et~al\mbox{.}(2022)]%
        {shandilya2022need}
\bibfield{author}{\bibinfo{person}{Esha Shandilya}, \bibinfo{person}{Mingming
  Fan}, {and} \bibinfo{person}{Garreth~W Tigwell}.}
  \bibinfo{year}{2022}\natexlab{}.
\newblock \showarticletitle{``I need to be professional until my new team uses
  emoji, GIFs, or memes first'': New Collaborators' Perspectives on Using
  Non-Textual Communication in Virtual Workspaces}. In
  \bibinfo{booktitle}{\emph{CHI '22}}. \bibinfo{pages}{1--13}.
\newblock


\bibitem[Sharma et~al\mbox{.}(2020)]%
        {sharma2017activation}
\bibfield{author}{\bibinfo{person}{Sagar Sharma}, \bibinfo{person}{Simone
  Sharma}, {and} \bibinfo{person}{Anidhya Athaiya}.}
  \bibinfo{year}{2020}\natexlab{}.
\newblock \showarticletitle{Activation functions in neural networks}.
\newblock \bibinfo{journal}{\emph{International Journal of Engineering Applied
  Sciences and Technology}} \bibinfo{volume}{6}, \bibinfo{number}{12}
  (\bibinfo{year}{2020}), \bibinfo{pages}{310--316}.
\newblock


\bibitem[Skovholt et~al\mbox{.}(2014)]%
        {skovholt2014communicative}
\bibfield{author}{\bibinfo{person}{Karianne Skovholt}, \bibinfo{person}{Anette
  Gr{\o}nning}, {and} \bibinfo{person}{Anne Kankaanranta}.}
  \bibinfo{year}{2014}\natexlab{}.
\newblock \showarticletitle{The communicative functions of emoticons in
  workplace e-mails::-}.
\newblock \bibinfo{journal}{\emph{Journal of Computer-Mediated Communication}}
  \bibinfo{volume}{19}, \bibinfo{number}{4} (\bibinfo{year}{2014}),
  \bibinfo{pages}{780--797}.
\newblock


\bibitem[Smith et~al\mbox{.}(2014)]%
        {smith2014social}
\bibfield{author}{\bibinfo{person}{Eliot~R Smith}, \bibinfo{person}{Diane~M
  Mackie}, {and} \bibinfo{person}{Heather~M Claypool}.}
  \bibinfo{year}{2014}\natexlab{}.
\newblock \bibinfo{booktitle}{\emph{Social psychology}}.
\newblock \bibinfo{publisher}{Psychology Press}.
\newblock


\bibitem[{Snap Inc.}(2022)]%
        {snap2022q1}
\bibfield{author}{\bibinfo{person}{{Snap Inc.}}}
  \bibinfo{year}{2022}\natexlab{}.
\newblock \showarticletitle{Snap Inc. announces first quarter 2022 financial
  results}.
\newblock  (\bibinfo{year}{2022}).
\newblock
\urldef\tempurl%
\url{https://investor.snap.com/news/news-details/2022/Snap-Inc.-Announces-First-Quarter-2022-Financial-Results}
\showURL{%
Retrieved Aug 22, 2022 from \tempurl}


\bibitem[Squartini et~al\mbox{.}(2013)]%
        {squartini2013reciprocity}
\bibfield{author}{\bibinfo{person}{Tiziano Squartini},
  \bibinfo{person}{Francesco Picciolo}, \bibinfo{person}{Franco Ruzzenenti},
  {and} \bibinfo{person}{Diego Garlaschelli}.} \bibinfo{year}{2013}\natexlab{}.
\newblock \showarticletitle{Reciprocity of weighted networks}.
\newblock \bibinfo{journal}{\emph{Scientific Reports}} \bibinfo{volume}{3},
  \bibinfo{number}{1} (\bibinfo{year}{2013}), \bibinfo{pages}{1--9}.
\newblock


\bibitem[Stuart(2010)]%
        {stuart2010matching}
\bibfield{author}{\bibinfo{person}{Elizabeth~A Stuart}.}
  \bibinfo{year}{2010}\natexlab{}.
\newblock \showarticletitle{Matching methods for causal inference: A review and
  a look forward}.
\newblock \bibinfo{journal}{\emph{Statistical Science: A Review Journal of the
  Institute of Mathematical Statistics}} \bibinfo{volume}{25},
  \bibinfo{number}{1} (\bibinfo{year}{2010}), \bibinfo{pages}{1}.
\newblock


\bibitem[Tang and Hew(2019)]%
        {tang2019emoticon}
\bibfield{author}{\bibinfo{person}{Ying Tang} {and} \bibinfo{person}{Khe~Foon
  Hew}.} \bibinfo{year}{2019}\natexlab{}.
\newblock \showarticletitle{Emoticon, emoji, and sticker use in
  computer-mediated communication: A review of theories and research findings}.
\newblock \bibinfo{journal}{\emph{International Journal of Communication}}
  \bibinfo{volume}{13} (\bibinfo{year}{2019}), \bibinfo{pages}{27}.
\newblock


\bibitem[Tigwell and Flatla(2016)]%
        {tigwell2016oh}
\bibfield{author}{\bibinfo{person}{Garreth~W Tigwell} {and}
  \bibinfo{person}{David~R Flatla}.} \bibinfo{year}{2016}\natexlab{}.
\newblock \showarticletitle{Oh that's what you meant! Reducing emoji
  misunderstanding}. In \bibinfo{booktitle}{\emph{Proceedings of the 18th
  International Conference on Human-Computer Interaction with Mobile Devices
  and Services Adjunct}}. \bibinfo{pages}{859--866}.
\newblock


\bibitem[Tsvetkova and Macy(2014)]%
        {tsvetkova2014social}
\bibfield{author}{\bibinfo{person}{Milena Tsvetkova} {and}
  \bibinfo{person}{Michael~W Macy}.} \bibinfo{year}{2014}\natexlab{}.
\newblock \showarticletitle{The social contagion of generosity}.
\newblock \bibinfo{journal}{\emph{PloS One}} \bibinfo{volume}{9},
  \bibinfo{number}{2} (\bibinfo{year}{2014}), \bibinfo{pages}{e87275}.
\newblock
\urldef\tempurl%
\url{https://doi.org/10.1371/journal.pone.0087275}
\showDOI{\tempurl}


\bibitem[Tung and Deng(2007)]%
        {tung2007increasing}
\bibfield{author}{\bibinfo{person}{Fang-Wu Tung} {and} \bibinfo{person}{Yi-Shin
  Deng}.} \bibinfo{year}{2007}\natexlab{}.
\newblock \showarticletitle{Increasing social presence of social actors in
  e-learning environments: Effects of dynamic and static emoticons on
  children}.
\newblock \bibinfo{journal}{\emph{Displays}} \bibinfo{volume}{28},
  \bibinfo{number}{4-5} (\bibinfo{year}{2007}), \bibinfo{pages}{174--180}.
\newblock


\bibitem[Utz(2000)]%
        {utz2000social}
\bibfield{author}{\bibinfo{person}{Sonja Utz}.}
  \bibinfo{year}{2000}\natexlab{}.
\newblock \showarticletitle{Social information processing in MUDs: The
  development of friendships in virtual worlds}.
\newblock \bibinfo{journal}{\emph{Journal of Online Behavior}}
  \bibinfo{volume}{1}, \bibinfo{number}{1} (\bibinfo{year}{2000}),
  \bibinfo{pages}{2002}.
\newblock


\bibitem[Wheeler(1966)]%
        {wheeler1966toward}
\bibfield{author}{\bibinfo{person}{Ladd Wheeler}.}
  \bibinfo{year}{1966}\natexlab{}.
\newblock \showarticletitle{Toward a theory of behavioral contagion.}
\newblock \bibinfo{journal}{\emph{Psychological Review}} \bibinfo{volume}{73},
  \bibinfo{number}{2} (\bibinfo{year}{1966}), \bibinfo{pages}{179}.
\newblock


\end{thebibliography}

\appendix

\end{document}